\documentclass[pra,twocolumn,showpacs,amsmath,amssymb]{revtex4}

\usepackage{graphicx}
\usepackage{epsfig}
\usepackage{amsmath}
\usepackage{ascmac} 
\usepackage{graphics} 
\usepackage{ascmac}
\usepackage{color}
\usepackage{amsmath,amssymb}
\usepackage{feynmp}
\usepackage{latexsym}
\begin{document}

\title{Stability Criterion for Superfluidity based on the Density Spectral Function}

\author{Shohei Watabe$^{1}$}  
\author{Yusuke Kato$^{2}$}
\affiliation{$^{1}$ 
Department of Physics, Faculty of Science, The University of Tokyo, Tokyo 113-0033, Japan
\\
$^{2}$ 
Department of Basic Science, The University of Tokyo, Tokyo 153-8902, Japan}

\begin{abstract} 
We study a stability criterion hypothesis for superfluids expressed in terms of the local density spectral function ${\mathcal I}_{n}({\bf r}, \omega)$ that is 
applicable to both homogeneous and inhomogeneous systems.  
We evaluate the local density spectral function 
in the presence of a one-dimensional repulsive/attractive external potential within Bogoliubov theory, using solutions for the tunneling problem. 
We also evaluate the local density spectral function using an orthogonal basis, 
and calculate the autocorrelation function $C_{n} ({\bf r},t)$. 
When superfluids in a $d$-dimensional system flow below a threshold, 
${\mathcal I}_{n}({\bf r}, \omega) \propto \omega^{d}$ holds in the low-energy regime 
and $C_{n} ({\bf r},t) \propto 1/t^{d+1}$ holds in the long-time regime. 
However, when superfluids flow with the critical current, 
${\mathcal I}_{n}({\bf r}, \omega) \propto \omega^{\beta}$ holds in the low-energy regime 
and $C_{n} ({\bf r},t) \propto 1/t^{\beta+1}$ holds in the long-time regime with $\beta < d$. 
These results support the stability criterion hypothesis recently proposed. 
\end{abstract}

\maketitle

\section{Introduction} 

The study of superfluids has revealed a cornucopia of fascinating phenomena as well as important concepts in the physics of condensed matter. 
Interesting phenomena related to superfluidity, such as phase slips and a persistent current, 
continue as topics of interest~\cite{Ramanathan2011,Wright2013} despite their long history of investigation. 
Since one notable feature is dissipationless flow below a threshold, 
the stability of superfluids is a very important issue. 
 
Although 
the Landau criterion provides the critical velocity and predicts that 
an ideal Bose gas is an unstable superfluid, 
many experimental results~\cite{Avenel1985, Varoquaux1986,Raman1999,Onofrio2000,Inouye2001,Engels2007} 
and numerical simulations~\cite{Frisch1992,Hakim1997,Heupe2000,Rica2001,Pham2002,Pavloff2002,Winiecke1999,Jackson2007,Aftalion2003} have shown that the critical velocity is actually smaller than Landau's critical velocity. 
(In cases where impurities are comparable in size to atoms, the critical velocity approaches Landau's critical velocity~\cite{Phillips1974, Ellis1980,Chikkatur2000}.) 
As is well known, the dissipation of superfluids at a smaller velocity than Landau's critical velocity 
is caused by emissions of phase defects, such as quantized vortices and solitons. 
Since the Landau criterion is based on the Galilean transformation, 
this criterion is applicable only to uniform systems. 
We thus need a stability condition for a superfluid flowing through an obstacle, in which case the translation invariance is broken. 

A feature of superfluids is the suppression of the density fluctuation. 
Although the compressibility diverges in an ideal Bose gas, it does not diverge in a Bose gas with a repulsive interaction. 
When we observe a two-body distribution function, the ideal Bose gas exhibits spatial density fluctuations and tends to form particle clusters due to the Bose statistics alone~\cite{Gruter1997}. 
On the other hand, 
a Bose gas with a repulsive interaction exhibits the ``density homogenization'' effect~\cite{Gruter1997} 
and its density fluctuations are suppressed in the long-wavelength regime~\cite{Kashurnikov2001}.  
In the Gross-Pitaevskii equation~\cite{Gross1961,Pitaevskii1961}, this homogenization effect may be included through the nonlinear effect on the macroscopic wave function~\cite{Hohenberg1965}. 
When the dissipation occurs in the superfluid above a threshold, emergent phase defects such as quantized vortices and solitons are often featured, 
but the density also fluctuates. In fact, the phase and density are canonical variables. 

Thus, we expect that the suppression of density fluctuations with respect to a perturbation characterizes the stability of superfluids. 
On the basis of this idea, we recently proposed a stability criterion hypothesis based on the local density spectral function ${\mathcal I}_{n}({\bf r}, \omega)$ or the autocorrelation function $C_{n} ({\bf r}, t)$~\cite{KatoWatabe2010JLTP,KatoWatabe2010PRL}. 
The former function is defined as 
\begin{align} 
{\mathcal I}_{n}({\bf r}, \omega) 
= & 
\sum\limits_{l} 
| \langle l| \delta \hat{n}({\bf r}) | {\rm g} \rangle |^{2} 
\delta (\omega - \omega_{l} + \omega_{\rm g}), 
\end{align} 
where $|{\rm g}\rangle$ is the ground state vector or a stable superflow state vector 
with the energy $\hbar \omega_{\rm g}$ and $\delta \hat{n}({\bf r})$ is the density fluctuation operator. 
($| l \rangle$ is a state vector of an excited state $l$ with the energy $\hbar \omega_{l}$.) 
The autocorrelation function is the Fourier transform of this function 
\begin{align}
C_{n} ({\bf r}, t) = & \int d \omega {\mathcal I}_{n} ({\bf r}, \omega) \cos (\omega t). 
\end{align} 
When a superflow current is $J \leq J_{\rm c}$, where $J_{\rm c}$ is the critical current, 
the local density spectral function in a $d$-dimensional system behaves as 
\begin{eqnarray}
\lim_{\omega\rightarrow 0}{\mathcal I}_{n}({\bf r}, \omega) \propto 
\left \{
\begin{array}{ll}
\omega^{\beta}  & \qquad (J=J_{\rm c})
\\
\omega^{d} &\qquad (J<J_{\rm c})
\end{array}
\right. 
\label{eq2}
\end{eqnarray} 
and the autocorrelation function behaves as 
\begin{eqnarray}
\lim_{t\rightarrow \infty} C_{n}({\bf r}, t) \propto  
\left \{
\begin{array}{ll}
1/t^{\beta+1}  & \qquad  (J=J_{\rm c})
\\
1/t^{d+1}  &\qquad (J<J_{\rm c}) 
\end{array}
\right.  
\label{eq3}
\end{eqnarray} 
with $\beta < d$. 
We have gathered only a few pieces of evidence for this criterion hypothesis~\cite{KatoWatabe2010JLTP,KatoWatabe2010PRL}. 

In this paper, 
we discuss the validity of the criterion 
by calculating the density spectral function 
not only for a one-dimensional repulsive potential barrier, but also for a one-dimensional attractive external potential 
using the tunneling solutions of Bogoliubov theory. 
In the latter case, the critical current $J_{\rm c}$ is equal to Landau's critical current. 
The density spectral function is enhanced at $J = J_{\rm c}$ in the low-energy regime 
far from the attractive potential; this is marked contrast to the case with the repulsive potential barrier. 
We also numerically demonstrate the validity of (\ref{eq3}) in the repulsive potential barrier case. 

We also discuss and numerically evaluate the density spectral function 
with the use of an orthogonal basis in the Bogoliubov approximation. 
An orthogonal basis is generally employed to calculate the spectral function, 
and tunneling solutions do not always satisfy the Bogoliubov orthonormalization condition. 
The tunneling solutions far from the potential barrier consist of the superposition of plane waves that satisfy the Bogoliubov normalization condition in the momentum space. 
Even if we use the orthogonal set, 
the low-energy behavior of the local density spectral function is qualitatively unchanged.

Section~\ref{SecIII} serves as an introduction to the local density spectral function. 
In Section~\ref{SecV}, we calculate the local density spectral function 
in the presence of the one-dimensional external potential. 
We calculate the density spectral function 
in a uniform system using Bogoliubov theory, and discuss the Landau instability in Section~\ref{SecVI}. 
Section~\ref{SecVII} also examines the density spectral function in Feynman's single-mode approximation and for an ideal Bose gas. 
Based on the results described in these sections, in Section~\ref{SecVIII}, 
we discuss the validity of the stability criterion hypothesis for superfluids in light of the density spectral function. 

We highlight results that were not addressed in the earlier short reports~\cite{KatoWatabe2010JLTP,KatoWatabe2010PRL}: 
(i) the comparative study of the local density spectral function for the repulsive/attractive potential barrier (Section~\ref{SecV}), 
(ii) the explicit formulas of the local density spectral function in the low-energy regime for the repulsive potential barrier case, 
obtained from the tunneling solutions at the critical current  (Section~\ref{SecV}), 
(iii) the spectral function calculated with an orthogonal basis, and a comparative study between this result and the spectral function obtained from the tunneling solutions (Section~\ref{SecV}), 
(iv) the numerically-calculated density spectral function in the uniform system using Bogoliubov theory (Section~\ref{SecVI}), 
(v) the application of the stability criterion hypothesis to an ideal Bose gas (Sections~\ref{SecVII} and~\ref{SecVIII}), 
and (vi) numerical evidence for the hypothesis in terms of the autocorrelation function (Section~\ref{SecVIII}).

\section{local density spectral function}\label{SecIII}

The density correlation function measured at ${\bf \rm x}_{1} = ({\bf  r}_{1},t_{1})$ and ${\bf \rm x}_{2} = ({\bf  r}_{2},t_{2})$ is provided by 
\begin{align}
C_{n} ({\bf \rm x}_{1}; {\bf \rm x}_{2}) 
= & 
\langle {\rm g} | \delta \hat n ({\bf \rm x}_{1})  \delta \hat n ({\bf \rm x}_{2}) | {\rm g} \rangle 
, 
\end{align} 
where 
$\delta \hat n ({\bf \rm x})$ is a density fluctuation operator 
\begin{align}
\delta \hat n ({\bf \rm x}) = \hat n({\bf \rm x}) - \langle {\rm g} | \hat n({\bf \rm x}) | {\rm g} \rangle, 
\end{align} 
and 
$| {\rm g} \rangle$ is a ket vector of the ground state or a stable superflow state of a Hamiltonian $\hat H$ 
satisfying $\hat H | {\rm g} \rangle = \hbar \omega_{\rm g} | {\rm g} \rangle$.  
Using the Fourier transformation, 
we obtain the spectral function 
\begin{align} 
& 
{\mathcal I}_{n} ({\bf r}_{1}, {\bf r}_{2}; \omega)
\nonumber 
\\
= & 
\frac{1}{2\pi}
\int_{-\infty}^{\infty} d (t_{2}-t_{1}) C_{n} ({\bf \rm x}_{1}; {\bf \rm x}_{2}) e^{- i \omega (t_{2} - t_{1})}
\nonumber 
\\
= &  
\sum\limits_{l}
\langle {\rm g}  | \delta \hat{n}({\bf r}_{1})  | l \rangle  \langle l | \delta \hat{n}({\bf r}_{2}) | {\rm g} \rangle 
\delta (\omega - \omega_{l} + \omega _{{\rm g}}) . 
\label{IABr1r2omega}
\end{align} 
Here,  $| l \rangle$ is a ket vector of an excited state with an index $l$ of the Hamiltonian $\hat H$ 
satisfying $\hat H | l \rangle = \hbar \omega_{l} | l \rangle$ with $\hbar \omega_{l} > \hbar \omega_{\rm g}$.

The local density spectral function ${\mathcal I}_{n}({\bf r}, \omega)$ 
and the autocorrelation function $C_{n} ({\bf r}, t)$ are local functions at ${\bf r} = {\bf r}_{1} = {\bf r}_{2}$, 
given by 
\begin{align}
{\mathcal I}_{n}({\bf r}, \omega) 
=  & {\mathcal I}_{n}({\bf r},{\bf r}; \omega) 
\\
= & 
\sum\limits_{l} 
| \langle l| \delta \hat{n}({\bf r}) | {\rm g} \rangle |^{2} 
\delta (\omega - \omega_{l} + \omega_{\rm g}), 
\label{InrOmega}
\\
C_{n} ({\bf r}, t) = & C_{n}^{\rm S} ({\bf r}, t ; {\bf r}, 0) 
= 
\int d \omega {\mathcal I}_{n} ({\bf r}, \omega) \cos (\omega t), 
\label{AutocorrlationCnrt}
\end{align} 
where $C_{n}^{\rm S}$ is 
the symmetrized correlation function 
\begin{align}
& C_{n}^{\rm S} ({\bf \rm x}_{1}; {\bf \rm x}_{2}) 
\nonumber 
\\
= & 
\frac{1}{2} 
[ 
\langle {\rm g} | \delta \hat n ({\bf \rm x}_{1})  \delta \hat n ({\bf \rm x}_{2}) | {\rm g} \rangle 
+ 
\langle {\rm g} | \delta \hat n ({\bf \rm x}_{2})  \delta \hat n ({\bf \rm x}_{1}) | {\rm g} \rangle
]. 
\end{align}

In the uniform system, the local density spectral function is related to the Fourier transformation of the dynamic structure factor as 
\begin{align}
{\mathcal I}_{n} ({\bf r}_{1} ,{\bf r}_{2} ; \omega) = 
\int\frac{d{\bf q}}{(2\pi)^{d}} S({\bf q}, \omega) e^{i {\bf q} \cdot ({\bf r}_{1} - {\bf r}_{2}) } 
\end{align} 
for dimensionality $d$. 
In this case, 
the equal point local density spectral function does not have ${\bf r}$-dependence, 
and is given by 
\begin{align}
 {\mathcal I}_{n} (\omega) = {\mathcal I}_{n} ({\bf r} ,{\bf r} ; \omega) = \int\frac{d{\bf q}}{(2\pi)^{d}} S({\bf q}, \omega) . 
 \label{eq11}
\end{align}

When we consider the fluctuations in the Bogoliubov level,  
the density fluctuation operator $\delta \hat{n}({\bf r},t)$ and 
the phase fluctuation operator that satisfy the canonical commutation relation $[\delta \hat{n}({\bf r}') ,  \delta \hat{\theta}({\bf r})] 
= 
i\delta({\bf r}-{\bf r}')$ 
are given by  
\begin{align} 
\delta \hat{n}({\bf r},t) = & A({\bf r}) \sum\limits_{j} 
\left [ G_{j} ({\bf r})  e^{- i E_{j}t/\hbar} \hat{a}_{j} + G_{j}^{*} ({\bf r})  e^{i E_{j}^{*}t/\hbar} \hat{a}_{j}^{\dag} \right ], 
\label{TB2-23} 
\\ 
\delta \hat{\theta}({\bf r},t) = & \frac{1}{2iA({\bf r})} \sum\limits_{j} 
\left [ S_{j} ({\bf r}) e^{- i E_{j}t/\hbar} \hat{a}_{j} - S_{j}^{*} ({\bf r}) e^{i E_{j}^{*}t/\hbar} \hat{a}_{j}^{\dag} \right ], 
\label{TB2-24} 
\end{align} 
where $\hat{a}_{j}^{}$ is the annihilation operator of the Bogoliubov excitation. 
$A ({\bf r})$ is the amplitude of the condensate wave function $\Psi_{0}({\bf r}) = A({\bf r})e^{i\theta_{0} ({\bf r})} $ 
that satisfies the stationary Gross-Pitaevskii equation 
\begin{align}
\hat{\mathcal H}_{0}  \Psi_{0}({\bf r}) = 0, 
\label{TB2-4} 
\end{align}
where 
\begin{align}
\hat{\mathcal H}_{0} = - \frac{\hbar^{2}}{2m} \nabla^{2} + V_{\rm ext} ({\bf r}) 
-\mu + g |\Psi_{0}({\bf r})|^{2}. 
\label{TB2-5} 
\end{align} 
Here, $m$ is the atomic mass, $V_{\rm ext} ({\bf r})$ is the external potential, $\mu$ is the chemical potential, 
and $g$ is the interaction strength.

The functions $G_{j} ({\bf r})$ and $S_{j} ({\bf r})$ are 
given by 
\begin{align}
G_{j} ({\bf r}) = & u_{j}({\bf r})e^{-i\theta_{0} ({\bf r})} - v_{j}({\bf r})e^{i\theta_{0} ({\bf r})}, 
\label{TB2-15} 
\\
S_{j} ({\bf r}) = & u_{j}({\bf r})e^{-i\theta_{0} ({\bf r})} + v_{j}({\bf r})e^{i\theta_{0} ({\bf r})}, 
\label{TB2-14} 
\end{align} 
where $u_{j}({\bf r})$ and $v_{j}({\bf r})$ satisfy the Bogoliubov equation 
\begin{align}
\begin{pmatrix}
\hat{\mathcal H}_{0} + g |\Psi_{0} |^{2} & - g \Psi_{0}^{2} 
\\
 g \left[  \Psi_{0}^{*} \right ]^{2} & - \hat{\mathcal H}_{0} - g |\Psi_{0} |^{2} 
\end{pmatrix}
\begin{pmatrix}
u_{j} \\ v_{j} 
\end{pmatrix}
= 
E_{j} 
\begin{pmatrix}
u_{j} \\ v_{j} 
\end{pmatrix}. 
\label{TB2-6} 
\end{align} 
The orthonormalization condition in Bogoliubov theory 
is 
\begin{align}
\int d {\bf r} [ u_{i}^{*} ({\bf r}) u_{j} ({\bf r}) - v_{i}^{*} ({\bf r}) v_{j} ({\bf r}) ]= & \delta_{ij}. 
\label{orthogonalCondi}
\end{align} 
This relation holds when $E_{i} \neq E_{j}^{*}$. 
In Bogoliubov theory, the local density spectral function can be reduced to 
\begin{align} 
{\mathcal I}_{n} ({\bf r}, \omega) = 
n_{0}({\bf r}) \sum\limits_{l} |G_{l} ({\bf r})|^{2} \delta (\omega - E_{l}/\hbar) , 
\end{align} 
where the condensate density is given by 
\begin{align}
n_{0} ({\bf r}) = A^{2} ({\bf r}). 
\end{align}
The density and phase operators are discussed in \cite{Shevchenko1992} for $\theta_{0} ({\bf r}) = 0$. 
Both (\ref{TB2-23}) and (\ref{TB2-24}) are extensions of these operators 
to the current carrying state case. 
Relations between these fluctuations and $(S,G)$, which are non-quantized versions, are discussed in~\cite{Takahashi2009,Takahashi2010}.

The energy and the length are scaled respectively by the Hartree energy 
$g n_{0}$ and the healing length $\xi =  \hbar/\sqrt{mgn_{0}}$, where $n_{0}$ is the condensate density in a uniform regime. 
The current density ${\bf J}$ is scaled by Landau's critical current $J_{0} = c_{\rm s} m n_{0}$. 
Here, $c_{\rm s}$ is the speed of the Bogoliubov phonon $c_{\rm s} = \sqrt{g n_{0}/m}$, which scales the fluid velocity ${\bf v} = \hbar \nabla \theta_{0} ({\bf r})/m$. 
We use $\overline{\bf r} =  {\bf r}/\xi$, 
$\overline{\nabla} =  \xi\nabla$, 
$\overline{\Psi}_{0}(\overline{{\bf r}}) =  \Psi_{0}({\bf r})/\sqrt{n_{0}}$, 
$\overline{V}_{\rm ext}(\overline{\bf r}) =  V_{\rm ext}({\bf r})/(g n_{0})$, 
$\overline{E} =  E/(g n_{0})$, $\overline{\bf J} = {\bf J}/J_{0}$, and $\overline{\bf v} = {\bf v}/c_{\rm s}$. 
For simplicity, we omit the bar below.

\section{local density spectral function in Bogoliubov theory}\label{SecV}

We discuss a stationary superfluid state in the presence of a one-dimensional external potential. 
The external potential has $x$-dependence and the translational invariance holds in the $y$- and $z$-directions. 
The superfluid flows along the $x$-direction, 
i.e., the current density ${\bf J}$ in the $y$- and $z$-directions is absent $(J_{y} = J_{z} = 0)$. 
In this case, the Gross-Pitaevskii equation can be reduced to~\cite{Baratoff1970,Pavloff2002,Seaman2005,Danshita2006}
\begin{align}
\hat{\mathcal H} A(x) =  0, 
\qquad 
A^{2}(x) \frac{d \theta_{0} (x)}{dx} =  J,
\label{TB2-9} 
\end{align} 
where 
\begin{align}
\hat{\mathcal H} = - \frac{1}{2} \frac{d^{2}}{dx^{2}} + \frac{J^{2}}{2 A^{4}(x)} + V_{\rm ext}(x) - \mu + A^{2} (x). 
\label{TB2-11} 
\end{align} 

An external potential $V_{\rm ext} (x)$ is localized around $x = 0$, i.e., $V_{\rm ext} (|x|\rightarrow \infty) = 0$. 
We solve the first equation in (\ref{TB2-9}) with the boundary conditions $A (x) = 1$ and $d A(x) / dx = 0$ at $x = \pm \infty$. 
The Gross-Pitaevskii equation at $|x| = \infty$ gives 
$\mu = 1 + J^{2} / 2$. 
According to the second equation in (\ref{TB2-9}),  the phase $\theta_{0}(x)$ 
and the phase difference $\varphi$~\cite{Baratoff1970} are given by 
\begin{align}
\theta_{0}(x) = & \theta_{0}(0)  + J x + J \int_{0}^{x}dx' \left ( \frac{1}{A^{2}(x')} - 1 \right ), 
\label{TB2-12} 
\\
\varphi = & J \int_{-\infty}^{\infty}dx \left ( \frac{1}{A^{2}(x)} - 1 \right ) . 
\label{TB2-13} 
\end{align}

\begin{figure}[tbp]
\begin{center}
\includegraphics[width=5cm]{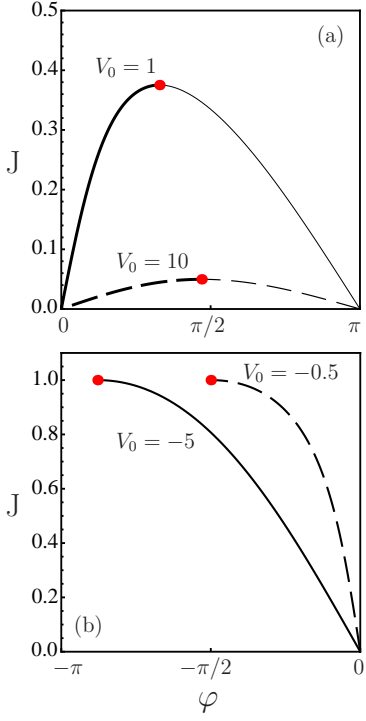}
\end{center}
\caption{
(Color online)
$J$-$\varphi$ relation. 
The delta-function potential $V_{\rm ext} (x) = V_{0} \delta(x)$ is used. 
Red points represent the critical current $J_{\rm c}$. 
(a) A repulsive potential case. 
Thick and thin lines are stable and unstable solutions, respectively, 
according to saddle-node bifurcation theory~\cite{Hakim1997}, and they merge at the critical current $J_{\rm c}$.   
(b) An attractive potential case. 
The current $J$ is an odd-function of the phase difference $\varphi$. 
The vertical axes in (a) and (b) are scaled by Landau's critical current $J_{0} = c_{\rm s}mn_{0}$.
$V_{0}$ is scaled by $g n_{0}/\xi$. 
}
\label{Fig1.fig}
\end{figure}

The current can flow without dissipation, when the phase is twisted ($\varphi \neq 0$) and the current is below the critical current $J_{\rm c}$. 
In the repulsive barrier case, stable branches (thick lines) and unstable branches (thin lines) 
merge at the maximum value of the stable supercurrent $J_{\rm c}$ 
with $d J / d \varphi = 0$ (Figure~\ref{Fig1.fig}(a)). 
The value $J_{\rm c}$ is less than the critical current of Landau's criterion $J=1$. 
This current phase relation can be also seen in Refs.~\cite{Baratoff1970,Sols1994,Watanabe2009,Piazza2010}. 
On the other hand, 
in an attractive potential case, 
the critical current $J_{\rm c}$ is always equal to Landau's critical current $J_{\rm c} = 1$ (Figure~\ref{Fig1.fig}(b))~\cite{Pavloff2002}. 
(To illustrate the current-phase relation in Figure~\ref{Fig1.fig}, we used the $\delta$-function potential barrier.) 

A {\it local} Landau criterion is occasionally quoted as the criterion giving the dissipation threshold in an inhomogeneous system 
that is less than the value in Landau's criterion. 
In this instability, 
excitations could be emitted if the velocity of the superfluid exceeded a threshold determined by the local density. 
In Bogoliubov theory, Landau's critical velocity 
is given by the speed of the Bogoliubov excitation $c_{\rm s}$. 
According to the local Landau's criterion, 
superfluidity would break at the position 
where the fluid speed $v ({\bf r})$ satisfies $v ({\bf r}) > c_{\rm s} ({\bf r}) \equiv \sqrt{n_{0} ({\bf r})}$. 

\begin{figure}[tbp]
\begin{center}
\includegraphics[width=6cm]{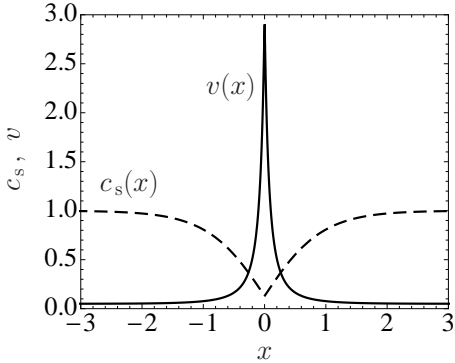}
\end{center}
\caption{
Velocity of superfluid $v(x) $ (solid line) 
and local speed of the Bogoliubov phonon $c_{\rm s} (x) = \sqrt{gn_{0}(x)/m}$ (dashed line), 
where the superfluid passes through the delta-function potential barrier 
$V_{\rm ext} (x) = V_{0} \delta(x)$ without dissipation. 
This result is obtained from the Gross-Pitaevskii equation. 
The current $J = 0.05$ is used, 
where the critical current in this case ($V_{0} = 7$ is taken) is $J_{\rm c} = 0.0707\cdots$. 
The vertical axis is scaled by the speed of the Bogoliubov phonon in the uniform system $c_{\rm s} = \sqrt{gn_{0}/m}$. The horizontal axis is scaled by the healing length $\xi$. $V_{0}$ is scaled by $g n_{0}/\xi$. } 
\label{Fig2.fig}
\end{figure}

This statement is not correct, however. 
Landau's criterion is applicable to the uniform system 
because it is based on a Galilean transformation. 
Furthermore, even if the speed of the fluid $v({\bf r})$ is larger than $c_{\rm s} ({\bf r})$,  
the state is stable. 
Indeed, in the stable superfluid state $J < J_{\rm c}$, we find $v (x) > c_{\rm s} (x)$ (Figure~\ref{Fig2.fig}). 
(In Figure~\ref{Fig2.fig}, we employed the $\delta$-function potential barrier.) 
According to the local Landau's criterion, 
this state is wrongly regarded as an unstable state. 
The local Landau's criterion works well only in the system locally homogeneous inside the barrier~\cite{Watanabe2009,Winiecki2000,Leszczyszyn2009,Piazza2013}. 

In the one-dimensional potential barrier case, 
the local density spectral function in the $d$-dimensional system is given by 
\begin{align}
{\mathcal I}_{n} (x, \omega) 
= & 
n_{0}(x)\int \frac{d{\bf k}^{\rm in }}{(2\pi)^{d}} |G (x; {\bf k}^{\rm in}) |^{2} 
\delta \left (
\omega - E (J, |{\bf k}^{\rm in} |, \theta) 
\right  ).  
\label{EQ28}
\end{align}  
In the tunneling problem, 
the incident momentum ${\bf k}^{\rm in} = (k_{x}^{\rm in}, k_{y}, k_{z})$ characterizes a state. 
The energy $E$ obtained from the Bogoliubov equation is 
\begin{align}
E (J, |{\bf k}|, \theta) = |{\bf k}| J \cos{\theta} + \sqrt{\frac{{\bf k} ^{2}}{2} \left ( \frac{ {\bf k}^{2}}{2} + 2 \right ) }, 
\label{TB2-26} 
\end{align} 
where 
$\theta$ is the angle between the wave vector ${\bf k} $ and the direction of the supercurrent density ${\bf J}$.

The wave function in the tunneling problem is given by 
\begin{align}
\begin{pmatrix}
u \\ v
\end{pmatrix} 
= & 
\tilde {\bf u}_{\mp} (x, k_{x}^{(1)}) 
+ 
r \tilde {\bf u}_{\mp} (x, k_{x}^{(2)}) 
+ 
a \tilde {\bf u}_{\mp} (x, k_{x}^{\mp})
\quad (x \rightarrow \mp \infty), 
\label{boundaryMP}
\\
\begin{pmatrix}
u \\ v
\end{pmatrix}
=  & 
t  \tilde {\bf u}_{\pm} (x, k_{x}^{(1)}) 
+ 
b \tilde {\bf u}_{\pm} (x, k_{x}^{\pm}) 
\quad (x \rightarrow \pm \infty), 
\label{boundaryPM}
\end{align} 
where 
\begin{align}
\tilde {\bf u}_{\pm} (x,k_{}) \equiv 
\begin{pmatrix}
\tilde u (k_{}) e^{+ i (Jx \pm \varphi /2)}
\\ 
\tilde v (k_{}) e^{- i (Jx \pm \varphi /2)}
\end{pmatrix} 
e^{ik_{}x}, 
\end{align} 
with 
\begin{align}
\begin{pmatrix}
\tilde u (k_{}) \\ \tilde v (k_{})
\end{pmatrix}
= 
{\cal N}^{-1} 
\begin{pmatrix}
1 \\
\displaystyle{
- E + \left ( \frac{k_{}^{2}}{2} + k_{} J + \frac{k_{\perp}^{2}}{2} +1 \right ) 
}
\label{NomalizedUV}
\end{pmatrix} . 
\end{align} 
In fact, the solution of the Bogoliubov equation in the uniform system is 
given by 
\begin{align}
\begin{pmatrix}
u \\ v
\end{pmatrix}
= 
e^{ (i k_{y} y + i k_{ z} z)} e^{i k_{x}^{ }x} 
\begin{pmatrix}
\tilde{u} (k_{x}) e^{+i[J x + {\rm sgn}(x) \varphi/2]}
\\
\tilde{v} (k_{x}) e^{-i[J x + {\rm sgn}(x) \varphi/2]}
\end{pmatrix} . 
\end{align} 
${\cal N}$ is the normalization coefficient determined from 
$|\tilde u|^{2} - |\tilde v|^{2} = 1$. 
$t$ and $r$ are the amplitude transmission and reflection coefficients, respectively. 
$k_{x}^{(1),(2),\pm}$ are 
the four solutions of 
\begin{align}
k_{x}^{4} + ( 2 k_{\perp}^{2} + 4 - 4 J^{2} ) k_{x}^{2} + 8 E J k_{x} & 
\nonumber 
\\ 
+ k_{\perp}^{4} + 4 k_{\perp}^{2} - 4 E^{2} & = 0, 
\label{TB2-28} 
\end{align} 
with respect to $k_{x}$, 
which comes from a dispersion relation 
\begin{align}
E = J k_{x} + \sqrt{\frac{k_{x}^{2} + k_{\perp}^{2}}{2} 
\left ( \frac{k_{x}^{2} + k_{\perp}^{2}}{2}  + 2 \right ) }, 
\label{EQ39}
\end{align}
where 
$k_{\perp} = \sqrt{k_{y}^{2} + k_{z}^{2}}$. 
$k_{x}^{(1)}$ is a real solution satisfying $k_{x}^{(1)} = k_{x}^{\rm in}$, and 
$k_{x}^{(2)}$ is the other real solution. 
The $k_{x}^{\pm}$ satisfy ${\rm sgn} ( {\rm Im} (k_{x}^{\pm}) ) = \pm 1$. 
The coefficients $t$, $r$, $a$, and $b$ are determined by solving 
(\ref{TB2-6}) with the boundary conditions (\ref{boundaryMP}) and (\ref{boundaryPM}). 
Details of the tunneling problem of the Bogoliubov excitation are summarized in Appendix~\ref{Appendix0}. 

\begin{figure}[tbp]
\begin{center}
\includegraphics[width=6cm]{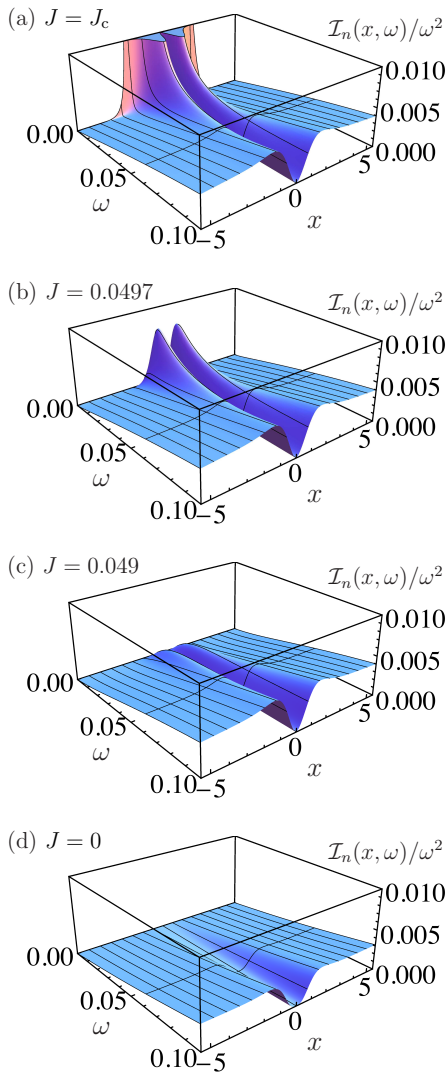}
\end{center}
\caption{
(Color online) Local density spectral function ${\mathcal I}_{n} (x,\omega)$ 
as functions of $\omega$ and $x$ in the three-dimensional case, 
in the presence of a repulsive delta-function potential barrier $V_{\rm ext} (x) = V_{0} \delta(x)$ with $V_{0} = 10$. 
In this case, the critical current $J_{\rm c}$ is $J_{\rm c} = 0.049753\cdots$. 
Here, ${\mathcal I}_{n} (x,\omega)$, $\omega$, $x$, and $J$ are scaled by 
$\hbar n_{0}/g$, $gn_{0}/\hbar$, $\xi$, and $J_{0}$, respectively. $V_{0}$ is scaled by $g n_{0}/\xi$. 
} 
\label{Fig3.fig}
\end{figure} 

\begin{figure}[tbp]
\begin{center}
\includegraphics[width=6cm]{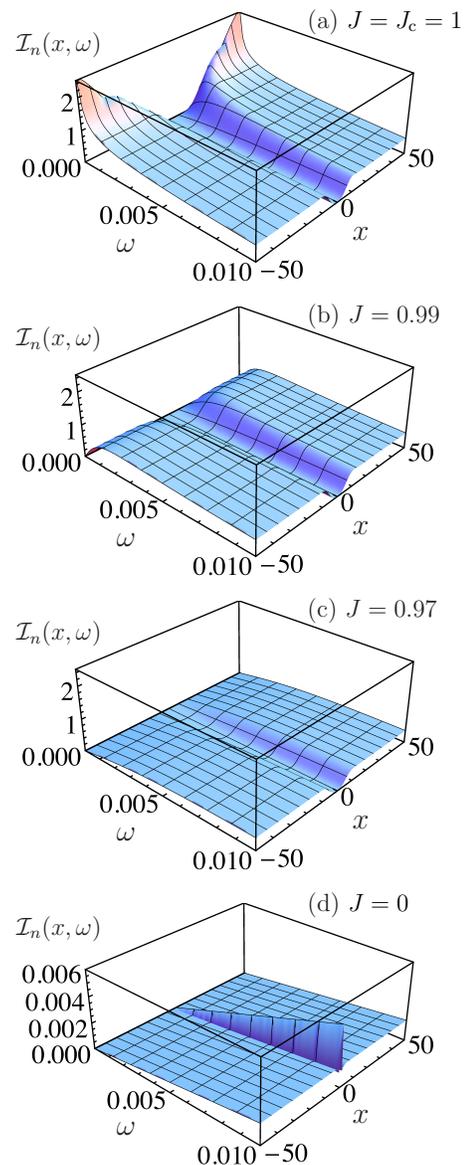}
\end{center}
\caption{
(Color online) Local density spectral function ${\mathcal I}_{n} (x,\omega)$ 
as functions of $\omega$ and $x$ in the one-dimensional case, 
in the presence of an attractive delta-function potential barrier $V_{\rm ext} (x) = V_{0} \delta(x)$ with $V_{0} = -2$. 
In this case, the critical current $J_{\rm c}$ is equal to Landau's critical current $J_{\rm c} = 1$. 
Here, ${\mathcal I}_{n} (x,\omega)$, $\omega$, $x$, and $J$ are scaled by 
$\hbar n_{0}/g$, $gn_{0}/\hbar$, $\xi$, and $J_{0}$, respectively. $V_{0}$ is scaled by $g n_{0}/\xi$. 
} 
\label{Fig4.fig}
\end{figure}

When the superfluid flows through the barrier, 
an anomaly of the density spectral function emerges around the region where the density is a minimum 
(Figures~\ref{Fig3.fig} and~\ref{Fig4.fig}). 
In the repulsive potential barrier case (Figure~\ref{Fig3.fig}), 
the enhancement of the local density spectral function appears around the barrier, 
which is located at $x=0$. 
This enhancement arises as the current $J$ approaches $J_{\rm c}$. 
(In Figure~\ref{Fig3.fig}, we used the $\delta$-function potential barrier. We have numerically checked the same behavior in the Gaussian-shaped potential barrier case.) 
For the attractive potential barrier (Figure~\ref{Fig4.fig}, where we also used the $\delta$-function potential barrier), 
the enhancement of the local density spectral function also occurs as the current $J$ approaches $J_{\rm c}$. 
However, it is located in a different region. 
The enhancement appears far from the attractive external potential, 
where the density is at a minimum and is also uniform.

\begin{figure}[tbp]
\begin{center}
\includegraphics[width=5cm]{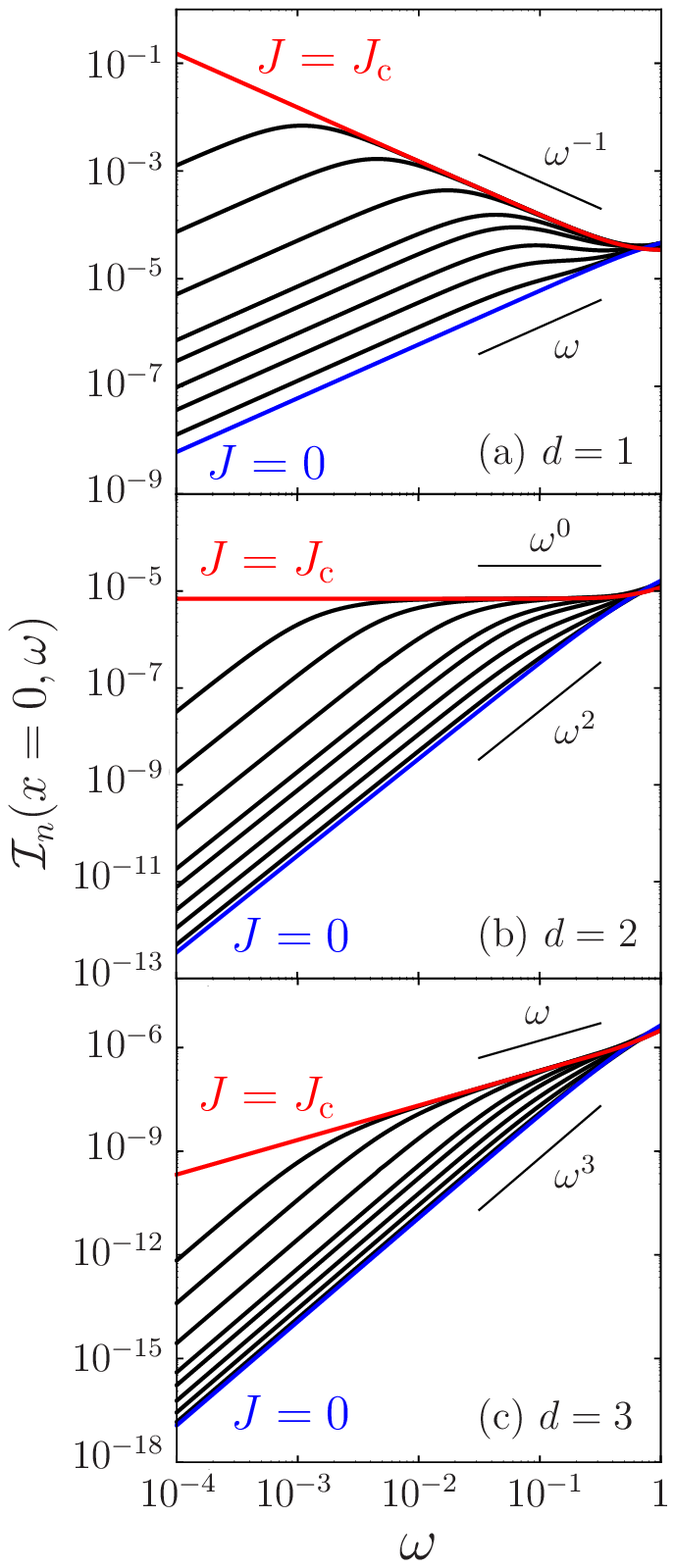}
\end{center}
\caption{
(Color online)
Local density spectral function ${\mathcal I}_{n} (x,\omega)$ at $x = 0$, 
in the presence of a repulsive delta-function potential barrier $V_{\rm ext} (x) = V_{0} \delta(x)$ with $V_{0} = 10$. 
(a), (b) and (c) are for the one-, two-, and three-dimensional systems, respectively. 
We used the set of the current 
$J = J_{\rm c} (= 0.049753\cdots), 0.04975, 0.0497, 0.049, 0.045, 0.04, 0.03, 0.02, 0.01,$ and $0$. 
Red and blue lines are respectively for $J = J_{\rm c}$ and $J = 0$. 
The functions are shifted from $J = 0$ to $J = J_{\rm c}$ with an increase in the current $J$. 
The vertical and horizontal axes are scaled by $\hbar n_{0}/g$ and $gn_{0}/\hbar$, respectively. 
The current $J$ is scaled by Landau's critical current $J_{0}$. $V_{0}$ is scaled by $g n_{0}/\xi$. 
} 
\label{Fig5.fig}
\end{figure}

The exponent of the local density spectral function in the low-energy regime in the state at $J = J_{\rm c}$ 
differs from the other states at $J < J_{\rm c}$ (Figure~\ref{Fig5.fig}). 
In a $d$-dimensional system at $J < J_{\rm c}$, the relation ${\mathcal I}_{n} (x,\omega)\propto \omega^{d}$ holds. 
At $J = J_{\rm c}$, on the other hand, the relation ${\mathcal I}_{n} (x,\omega)\propto \omega^{d-2}$ holds. 
(In Figure~\ref{Fig5.fig}, we used the repulsive $\delta$-function potential barrier. We have numerically checked the same exponent with respect to the $\omega$-dependence in the Gaussian-shaped potential barrier case.) 
The anomaly of the local density spectral function for an attractive potential case 
originates essentially from the Landau instability. 
The exponent of the local density spectral function 
will be discussed in Section~\ref{SecVI}.

\begin{figure}[tbp]
\begin{center}
\includegraphics[width=6cm]{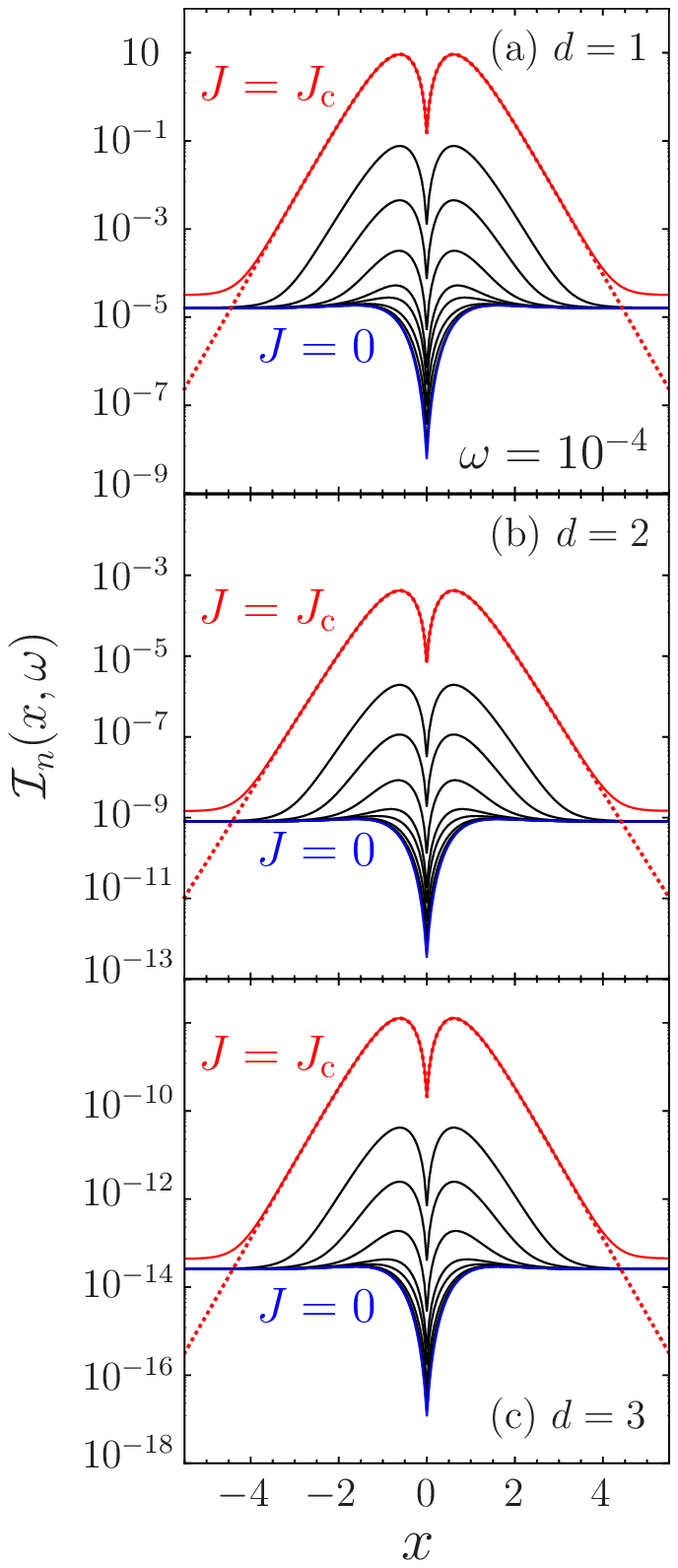}
\end{center}
\caption{
(Color online)
The spatial dependence of the local density spectral function ${\mathcal I}_{n} (x, \omega)$ 
at $\omega = 10^{-4}$ 
in the presence of a repulsive delta-function potential barrier $V_{\rm ext} (x) = V_{0} \delta(x)$ with $V_{0} = 10$. 
(a), (b), and (c) are for the one-, two-, and three-dimensional systems, respectively. 
We used the set of the current 
$J = J_{\rm c} (= 0.049753\cdots), 0.04975, 0.0497,0.049,0.045,0.04,0.03,0.02,0.01$, and $0$. 
Red and blue lines are for $J = J_{\rm c}$ and $J = 0$, respectively. 
The functions are shifted from $J = 0$ to $J = J_{\rm c}$ as the current $J$ increases. 
Red dotted lines are analytical results from (\ref{Current5-33}). 
The vertical and horizontal axes are scaled by $\hbar n_{0}/g$ and $\xi$, respectively. 
The current $J$ is scaled by Landau's critical current $J_{0}$. $V_{0}$ is scaled by $g n_{0}/\xi$. 
} 
\label{Fig6.fig}
\end{figure}

At $J = J_{\rm c} (<1)$ in the repulsive potential case, 
we can derive an analytic form of the local density spectral function in the low-energy regime. 
For dimensionality $d$, we have 
\begin{align}
{\mathcal I}_{n} (\omega, x) \simeq 
& 
\frac{ {\mathcal F}_{d} }{\pi} \omega^{d-2}
\left [ 
\partial_{\varphi} n_{0}(x) 
\right ]^{2}  , 
\label{Current5-33}
\end{align} 
where 
\begin{eqnarray} 
{\mathcal F}_{d} = 
\left \{ 
\begin{array}{lll}
\displaystyle{
\frac{2 J_{\rm c}^{2} }{ J_{\rm c}^{2} + \eta^{2} } 
} 
& 
\displaystyle{
(d=1)
}
\\
\displaystyle{
1 - \frac{\eta}{ \sqrt{J_{\rm c}^{2} + \eta^{2}} } 
}
& 
\displaystyle{
(d=2)
}
\\
\displaystyle{
\frac{1}{\pi} \left [ 1 - \frac{\eta}{J_{\rm c}} \tan^{-1} \left ( \frac{J_{\rm c}}{\eta} \right ) \right ]  
} 
& 
\displaystyle{
(d=3) 
}
\end{array}
\right . 
\end{eqnarray} 
with 
\begin{align}
\displaystyle{
\eta= \frac{\int_{-\infty}^{\infty}dx A(x) A_{\varphi}(x) }{\int_{-\infty}^{\infty}dx A_{\varphi}(x)/A^{3}(x)}
} 
\end{align} 
and $A_{\varphi} (x) = \partial A(x) /\partial \varphi$. 
Derivations may be found in Appendix~\ref{Appendix4}. 
Here, the barrier was assumed to be strong, leading to $J_{\rm c} \ll 1$. 
We also assumed $|\eta| \ll 1$, because $\eta = {\mathcal O} (J)$ as discussed in Appendix~\ref{Appendix2}. 
The spatial dependence of the local density spectral function ${\mathcal I}_{n} (x, \omega)$ 
is consistent with our analytical result (\ref{Current5-33}) (Figure~\ref{Fig6.fig}). 
When $\omega$ decreases, our analytical and numerical results agree over the wider range of $x$.

In Figure~\ref{Fig6.fig}, we used the $\delta$-function potential barrier. We have numerically checked the agreement between 
the numerical results and our analytical result (\ref{Current5-33}) in the Gaussian-shaped potential barrier case. 
Equation (\ref{Current5-33}) is applied to the potential barrier with the general shape. 
In fact, to derive (\ref{Current5-33}), we employed the wave function obtained without assuming the specific shape of the potential barrier. 
(See Appendices~\ref{Appendix2} and~\ref{Appendix4}). 

The use of the tunneling solutions facilitates the evaluation of the local density spectral function at the thermodynamic limit. 
However, generally speaking, we should use an orthogonal set when we evaluate the spectral functions. 
As shown below, even if we use the orthogonal set, 
our main results for the low-energy behavior of the local density spectral function are unchanged. 

The local density spectral function in the $d$-dimensional system is reduced to 
\begin{align}
{\mathcal I}_{n} (x, \omega) = \sum\limits_{l} M(x, E_{l}) \frac{1}{L^{d}} \delta (\omega - E_{l}), 
\end{align}
where $M(x, E_{l})$ is the squared matrix element given by 
\begin{align}
M (x, E_{l}) = & L^{d} | u_{l} (x) \Psi_{0}^{*} (x) - v_{l} (x) \Psi_{0} (x) |^{2} 
\\
= & L^{d} n_{0}(x) | G_{l} (x)|^{2}. 
\end{align}
Here, $L$ is the system size. 
To obtain $M (x, E_{l})$, we solve (\ref{TB2-6}) 
with the periodic boundary conditions 
\begin{align}
u_{l}(L/2) = u_{l}(-L/2), & \quad \partial_{x} u_{l}(L/2) = \partial_{x} u_{l}(-L/2) , 
\label{PBu}
\\
v_{l}(L/2) = v_{l}(-L/2) , & \quad \partial_{x} v_{l}(L/2) = \partial_{x} v_{l}(-L/2), 
\label{PBv}
\end{align}
and the normalization condition 
\begin{align}
\int_{-L/2}^{L/2} dx [|u_{l}(x)|^{2} - |v_{l}(x)|^{2}] = 1. 
\label{NrmCnd}
\end{align} 

To determine the spectral function, 
a calculation is needed at the thermodynamic limit. 
Although it is difficult to solve the Bogoliubov equation numerically at this limit,  
we have analytic solutions for a one-dimensional system with the $\delta$-function potential barrier $V_{\rm ext}(x) = V_{0} \delta (x)$~\cite{Danshita2006}. 
The solution $ {\bf u}_{+} \equiv (u_{+}, v_{+})^{\rm T}$ at $x \geq 0$ 
(${\bf u}_{-} \equiv (u_{-}, v_{-})^{\rm T}$ at $x < 0$)
are now given by 
\begin{align}
{\bf u}_{\pm} (x) = & \sum\limits_{
\substack{
k = k_{x}^{(1)}, k_{x}^{(2)}, k_{x}^{+}, k_{x}^{-}} } c_{\pm,k} 
{\bf U}_{\pm} (x,k), 
\label{eq116}
\end{align}
where 
\begin{align}
{\bf U}_{\pm} (x,k) 
= 
\begin{pmatrix} 
\{ [ 1 + k^{2} /(2E) ] \gamma (x) 
\mp i K_{u} (x,k) \} 
e^{i [(k+J)x \pm \varphi/2]} 
\\ 
\{ 
[ 1 - k^{2} /(2E) ] \gamma (x) 
\pm i K_{v} (x,k)  
\} e^{i [(k-J)x \mp \varphi/2]} 
\end{pmatrix}, 
\label{eq47}
\end{align} 
with 
\begin{align}
K_{u,v} (x,k) 
= J + \frac{k}{2E} 
[1 - J^{2} - \gamma^{2} (x) ] 
+ \frac{k^{3}}{4E} 
\pm \frac{k}{2} 
. 
\label{eq49}
\end{align} 
For (\ref{eq49}), the upper (lower) sign is for $K_{u}$ ($K_{v}$). 
Here, $\gamma (x)$ is related to the amplitude of the condensate wave function 
$A(x) = \sqrt{J^{2} + \gamma^{2} (x)}$ given by 
\begin{align}
\gamma (x) = & 
\sqrt{1 - J^{2}} \tanh[\sqrt{1 - J^{2}} (|x| + x_{0})]. 
\end{align}
$x_{0}$ is determined from the boundary condition of $\Psi_{0} (x)$ at $x=0$~\cite{Danshita2006}.

We determine eight coefficients $c_{\pm,k}$ and eigenenergy $E_{l}$ using (\ref{PBu}), (\ref{PBv}), (\ref{NrmCnd}),
and the boundary conditions at $x=0$ given by 
\begin{align}
{\bf u}_{+} (0) = {\bf u}_{-} (0), & \quad \partial_{x} {\bf u}_{+} (0) - \partial_{x} {\bf u}_{-} (0) = 2 V_{0} {\bf u}_{+} (0), 
\label{eq50}
\end{align}
Since ${\bf U}_{\pm} (x,k)$ are solutions of the Bogoliubov equation, 
(\ref{eq116}) satisfies the orthogonality (\ref{orthogonalCondi}) when $E_{l} \neq E_{l'}$.  

\begin{figure}[tbp]
\begin{center}
\includegraphics[width=7cm]{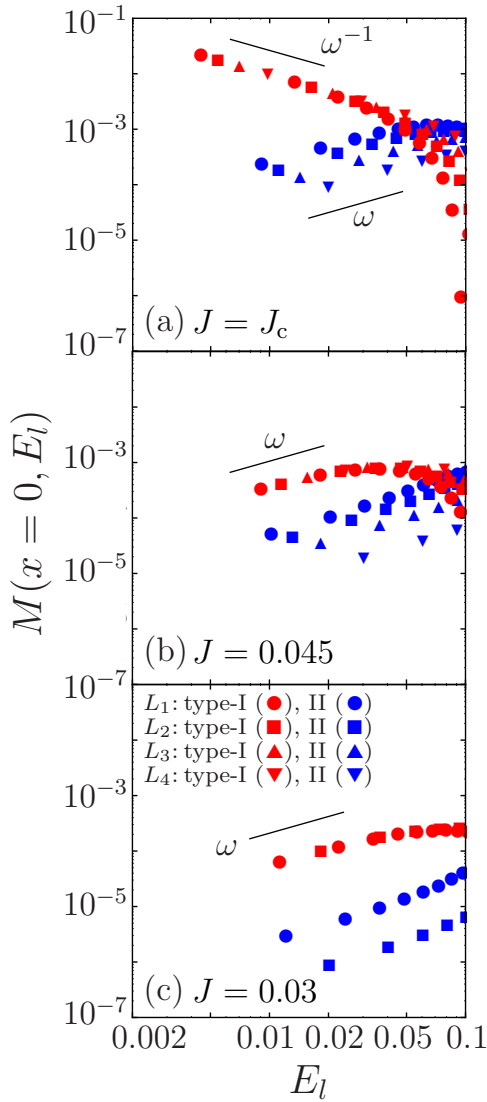}
\end{center}
\caption{(Color online) 
Squared matrix element $M(x,E_{l})$ of the local density spectral function at $x=0$ as a function of eigenenergy $E_{l}$ in the one-dimensional case. 
We used the barrier $V_{\rm ext} (x) = V_{0} \delta (x)$ with $V_{0} = 10$. 
(a) $J = J_{\rm c} = 0.049753\cdots$. (b) $J = 0.045$. (c) $J = 0.03$. 
The system sizes we used are (a) $(L_{1},L_{2},L_{3}, L_{4}) = (696,570,444,317)$, 
(b) $(L_{1}, L_{2}, L_{3}, L_{4}) = (630,490,351,211)$, 
and (c) $(L_{1}, L_{2}) = (525,316)$,  
where the decimal point is suppressed. These $L_{i}$ are determined from the periodic boundary conditions 
of the condensate wave function with a given $J$. 
The type-I is the excitation which makes a significant contribution to the matrix element at $J \neq 0$ for low $E_{l}$. 
The vertical and horizontal axes are scaled by $\xi n_{0}^{2}$ and $gn_{0}$, respectively. 
$J$, $\omega$, $L_{1,2,3,4}$ and $V_{0}$ are scaled by $J_{0}$, $gn_{0}/\hbar$, $\xi$, and $g n_{0}/\xi$, respectively. 
}
\label{Fig7.fig}
\end{figure}

The relation between the eigenenergy $E_{l}$ 
and the squared matrix element $M (x=0,E_{l})$ reveals two types of excitations (Figure~\ref{Fig7.fig}). 
The type-I excitation dominantly contributes the density fluctuations at $J \neq 0$, 
whose matrix element becomes larger for lower energies, in particular at $J = J_{\rm c}$. 
The contributions of the type-II excitation to the density fluctuations are smaller than those of type-I at $J \neq 0$, 
whose matrix element becomes smaller for lower energies at an arbitrary $J (\leq J_{\rm c})$. 
The first excitation is always type-I. 
The parity rule holds in the low-energy regime; the odd (even)-numbered excitations belong to type-I (II). 
In higher-energy regimes, it is difficult to distinguish between the two types of excitations. 
At $J = 0$, we cannot distinguish type-I from type-II because of degeneracy. 
(In Figure~\ref{Fig7.fig}, we used the $\delta$-function potential barrier.) 

When we plot the squared matrix element for several system sizes, 
the type-I excitation produces a smooth line in the low-energy regime (Figure~\ref{Fig7.fig}). 
We can thus introduce an interpolation function $\tilde M (x, \omega)$ 
satisfying two conditions; 
\begin{align}
\tilde M (x, E_{l}) = M (x, E_{l}), \quad (l \in  \textrm{type-I}), 
\end{align} 
and 
\begin{align}
|\partial \tilde M (x, \omega) / \partial \omega| \Delta E \ll |\tilde M (x, \omega)| . 
\end{align} 
$\tilde M (x, \omega)$ traces the squared matrix element $M (x, \omega)$ of the type-I excitation, 
and is a slowly-varying function of $\omega$ compared to the energy interval $\Delta E = |E_{l+2} - E_{l}|$, where $l \in$ type-I. 
In this expression, 
the type-I excitation is labeled with $l=1,3,5,\cdots$ 
in order of increasing $E_{l}$, using the parity rule.

Exponents of $\tilde M (x, \omega)$ (and also $ M (x, \omega)$ for the type-I excitation) with respect to $\omega$ are different 
between the cases at $J = J_{\rm c}$ and those at $J < J_{\rm c}$ (Figure~\ref{Fig7.fig}). 
These are $\omega^{-1}$ at $J = J_{\rm c}$ and $\omega$ at $J < J_{\rm c}$. 
In the stable superfluid state at $J < J_{\rm c}$, 
the zero-energy mode is only the phase mode, 
so that the low-energy solution is given by 
\begin{align} 
\begin{pmatrix}
S_{j}(x) \\  G_{j}(x)
\end{pmatrix}
= 
\frac{c}{\sqrt{E_{j}}}
\left [ 
\begin{pmatrix}
A(x) \\  0
\end{pmatrix}
+ E_{j}
\begin{pmatrix}
\tilde S (x) \\  \tilde G(x)
\end{pmatrix}
+ {\mathcal O}(E_{j}^{2})
\right ]. 
\label{eq123}
\end{align} 
Here, $c/\sqrt{E_{j}}$ is the normalization coefficient, and $\tilde S (x)$ and $\tilde G(x)$ are higher orders of $E_{j}$. 
At $J = J_{\rm c}$, however, 
the density mode related to $G (x)$ appears even at the zero-energy limit~\cite{Takahashi2009}, given by  
\begin{align}
G_{j}(x) = \frac{c_{\rm c}}{\sqrt{E_{j}}} \frac{\partial A (x) }{\partial \varphi} . 
\label{eq124}
\end{align} 
Details are provided in Appendix~\ref{Appendix2}. 
Here, $c_{\rm c}/\sqrt{E_{j}}$ is also the normalization coefficient. 
Using these solutions, we obtain the squared matrix element as 
\begin{eqnarray} 
\tilde M (x, \omega)  \simeq 
\left \{ 
\begin{array}{lll}
\displaystyle{
\omega n_{0}(x) |\tilde G (x)|^{2}
}  
& 
\displaystyle{
(J < J_{\rm c})
}
\\ 
\displaystyle{
\omega^{-1} 
\left [ 
\partial_{\varphi} n_{0}(x) 
\right ]^{2}
} 
& 
\displaystyle{
(J = J_{\rm c}), 
} 
\end{array}
\right . 
\label{eq125}
\end{eqnarray} 
at the low-energy regime up to a constant factor.

When we introduce the coarse-grained density of states 
\begin{align}
\tilde D_{d} (\omega) = \frac{1}{\delta} \int_{\omega - \delta /2}^{\omega + \delta /2} d \omega' 
\frac{1}{L^{d}} \sum\limits_{l} \delta (\omega ' - E_{l}), 
\end{align} 
the local density spectral function in the low-energy regime for $d=1$ is reduced to 
\begin{align}
{\mathcal I}_{n} (x, \omega) = \tilde M(x, \omega) \tilde D_{d=1} (\omega) . 
\label{eq127}
\end{align} 
Here, $\delta$ satisfies an arbitrarily small value satisfying $\Delta E \ll \delta \ll 1$ for large $L$. 
$\tilde D_{d} (\omega)$ is a smooth function, and we consider it to be the density of states at the thermodynamic limit. 
At this limit, 
we approximate $\tilde D_{d} (\omega)$ as 
\begin{align}
\tilde D_{d} (\omega) = & 
\int \frac{d{\bf k}^{\rm in }}{(2\pi)^{d}} 
\delta \left (
\omega - E (J, |{\bf k}^{\rm in}|, \theta)
\right  ) 
. 
\end{align} 
When $J < 1$, the excitation is a phonon, i.e., $|{\bf k}^{\rm in}| \propto \omega$, 
so that we obtain $\tilde D_{d} (\omega) \propto \omega^{d-1}$. 
As a result, for the dimensionality $d = 1$, ${\mathcal I}_{n}  \propto \omega$ holds at $J < J_{\rm c}$. 
At $J = J_{\rm c}$, ${\mathcal I}_{n} \propto \omega^{-1}$ holds.

For $d = 2$, we classify the eigenstates $l$ by $\theta \in [-\pi/2, \pi/2]$. 
We introduce infinitesimally small intervals $\Delta\theta_{m} \equiv [m \Delta \theta, (m+1) \Delta \theta]$ 
for $m \in [-N/2, (N/2)-1]$, where $1 \ll \pi / \Delta \theta \equiv N$ and $m \in \mathbb{Z}$ . 
In this case, the eigenstate can be labeled as $E_{l} = E_{l',m}$. 
The density spectral function is given by 
\begin{align}
{\mathcal I}_{n} (x, \omega) = \sum\limits_{l',m} M(x, E_{l',m}) \frac{1}{L^{2}} \delta (\omega - E_{l',m}). 
\end{align} 
We can discuss the case for $d=3$ in a similar way. 
Since $k_{\perp} = {\mathcal O}(E)$, 
the Bogoliubov equation with $k_{\perp}^{2}$ can be reduced to that for the one-dimensional case within ${\mathcal O} (E)$. 
In the low-energy regime, the solution has the same structure as (\ref{eq123}) at $J<J_{\rm c}$ 
or (\ref{eq124}) at $J = J_{\rm c}$. 
As a result, the $\omega$-dependence of the squared matrix element is also the same as (\ref{eq125}). 
The excitation is a phonon at $J < 1$, 
so that the $\omega$-dependence of the remaining factor of ${\mathcal I}_{n}$ is proportional to $\omega^{d-1}$. 
We thus end with 
\begin{align}
{\mathcal I}_{n} (x, \omega) \simeq & 
\left \{ 
\begin{array}{lll}
\displaystyle{
\omega^{d} n_{0} (x) |\tilde G (x)|^{2}
}  
& 
\displaystyle{
(J < J_{\rm c})
}
\\ 
\displaystyle{
\omega^{d-2} 
\left [ 
\partial_{\varphi} n_{0} (x) 
\right ]^{2}
} 
& 
\displaystyle{
(J = J_{\rm c}) 
} 
\end{array}
\right . 
\label{eq138}
\end{align}
at the low-energy regime up to a constant factor. 
This $\omega$-dependence is consistent with the results obtained from the tunneling solutions 
in the presence of the repulsive potential barrier.

\begin{figure}
\begin{center}
\includegraphics[width=7cm]{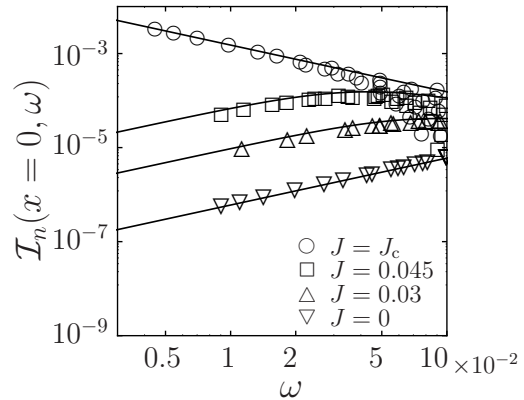}
\end{center}
\caption{
The local density spectral function at $x=0$ as a function of energy $\omega$. 
Each symbol represents $M(x=0,\omega) \tilde D_{1} (\omega)$ for 
$J = J_{\rm c} (= 0.049753\cdots)$ (circle), $0.045$ (square), 
$0.03$ (triangle), and $0$ (inverted-triangle) at $\omega = E_{l}$ for $l \in $ type-I. 
We used the barrier $V_{\rm ext} (x) = V_{0} \delta (x)$ with $V_{0} = 10$. 
These results are obtained from several system sizes. 
For $J = J_{\rm c}$, $0.045$, and $0.03$, we took the system sizes used in Fig.~\ref{Fig7.fig}. 
For $J=0$, we used the same system sizes $L_{1,2,3,4}$ as the case at $J = J_{\rm c}$. 
The solid lines show the local density spectral function produced from the solutions of the tunneling problem. 
The vertical and horizontal axes are scaled by $\hbar n_{0}/g$ and $g n_{0}/\hbar$, respectively. 
The current $J$ is scaled by Landau's critical current $J_{0}$. $V_{0}$ is scaled by $g n_{0}/\xi$. 
} 
\label{Fig8.fig}
\end{figure} 

In the low-energy regime, 
the local density spectral function ${\mathcal I}_{n}$ constructed from the tunneling solutions 
reproduces well the $\omega$-dependence of $M(x = 0, E_{l}) \tilde D_{d=1} (E_{l})$ for $l \in$ type-I (Figure~\ref{Fig8.fig}). 
(In Figure~\ref{Fig8.fig}, we used the $\delta$-function potential barrier.) 
On this basis, 
we can use the solutions of the tunneling problem to effectively calculate the local density spectral function at the thermodynamic limit, 
and to discuss the $\omega$-dependence of the local density spectral function at the low-energy limit.

\begin{figure}[tbp]
\begin{center}
\includegraphics[width=6cm]{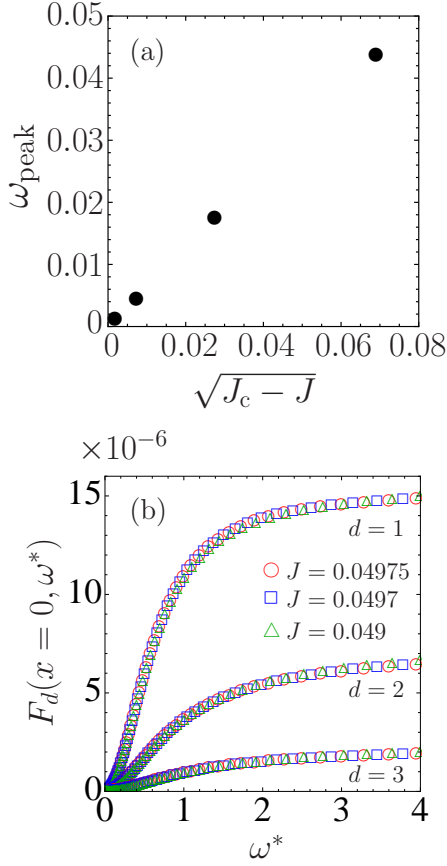}
\end{center}
\caption{
(Color online) 
(a) The frequency $\omega_{\rm peak}$ giving the peak of the local density spectral function ${\mathcal I}_{n} (x, \omega)$ at $x=0$ as a function of 
the scaling factor $\sqrt{J_{\rm c}-J}$ in the one-dimensional system. 
The data are taken from the result in Figure~\ref{Fig5.fig} (a). 
The vertical and horizontal axes are scaled by $g n_{0}/\hbar$ and $\sqrt{J_{0}}$, respectively. 
(b) The scaling function $F_{d} (x,\omega^{*}) = \omega^{2-d} {\mathcal I}_{n} (x, \omega)$ at $x=0$ as a function of 
the scaled energy (frequency) $\omega^{*} = \omega /\sqrt{J_{\rm c}-J}$, 
in one-, two-, and three-dimensional systems. 
Each symbol represents data at 
$J = 0.04975$ (circle), $0.0497$ (square) and $0.049$ (triangle). 
The result (b) is referred from~\cite{KatoWatabe2010PRL}. 
The vertical and horizontal axes are scaled by $(g/\hbar)^{1-d} n_{0}^{3-d}$ and $g n_{0}/(\hbar \sqrt{J_{0}})$, respectively. 
In both (a) and (b), we used the delta-function potential barrier $V_{\rm ext} (x) = V_{0} \delta(x)$ with $V_{0} = 10$, 
and its critical current is $J_{\rm c} = 0.049753\cdots$. 
Here, $V_{0}$ and $J$ is scaled by $gn_{0}/\xi$ and $J_{0}$, respectively. 
}
\label{Fig9.fig}
\end{figure} 

Hakim discussed the soliton instability as a saddle-node bifurcation, where 
the stable and unstable branches merge at the bifurcation point $J = J_{\rm c}$~\cite{Hakim1997}. 
Near the saddle node bifurcation point, a dynamical scaling relation can be found. 
An example of a dynamical scaling relation is the emission rate $\Gamma$ of the gray soliton given by 
$\Gamma \propto \sqrt{|V-V_{\rm c}|}$~\cite{Pham2002}. 
Here, $V$ is the strength of the potential barrier and $V_{\rm c}$ is its critical strength. 
The scaling law also holds between the scaling factor $\sqrt{J_{\rm c}-J}$ and 
the peak frequency $\omega_{\rm peak}$ that gives the peak of the local density spectral function at $x=0$ (Figure~\ref{Fig9.fig}(a)). 

The scaling function $F_{d} (x, \omega^{*} = \omega / \sqrt{J_{\rm c}-J})$ describes the universal behaviors of the local density spectral function near the critical current. 
For the dimensionality $d$, it is given by 
\begin{align}
{\mathcal I}_{n} (x, \omega ) = \omega^{d-2} F_{d} (x, \omega |J - J_{\rm c}|^{-1/2}). 
\end{align} 
In each dimension, 
the local density spectral functions near the critical current collapse onto a single curve, 
which implies a dynamical scaling law (Figure~\ref{Fig9.fig}(b)). 

These results in Figure~\ref{Fig9.fig} are obtained in the $\delta$-function potential barrier case. 
This dynamical scaling law may hold in the repulsive potential barrier case with the general shape and the arbitrary strength. 
In fact, this scaling law is a general property around the bifurcation point.

\section{Landau Instability in Bogoliubov theory}\label{SecVI}

\begin{figure}[tbp]
\begin{center}
\includegraphics[width=6cm]{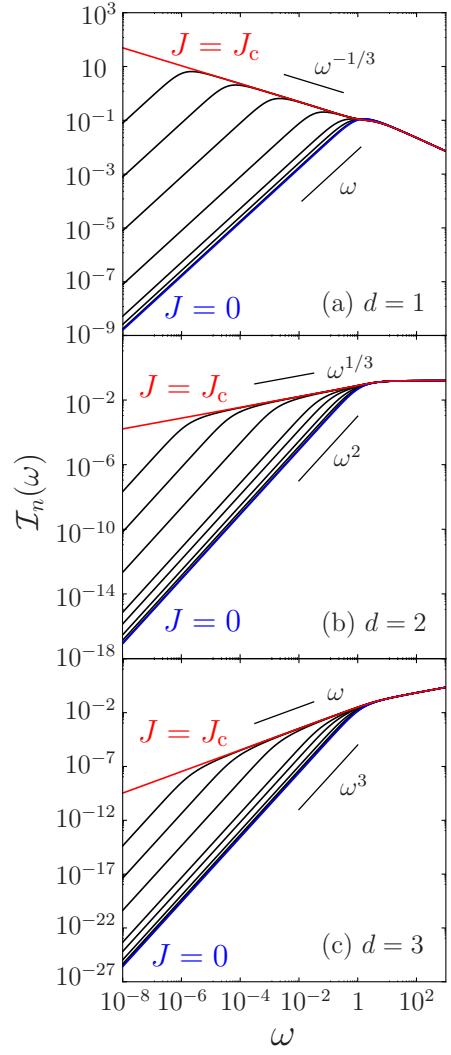}
\end{center}
\caption{
(Color online) 
Numerically-calculated density spectral function ${\mathcal I}_{n} (\omega)$ 
in the uniform system within Bogoliubov theory. 
(a), (b) and (c) are for the one-, two-, and three-dimensional systems, respectively. 
We used the set of the current 
$J = J_{\rm c}= 1, 0.999, 0.99, 0.9, 0.8, 0.6, 0.4, 0.2$, and $0$. 
Red and blue lines are for $J = J_{\rm c}$ and $J = 0$, respectvely. 
The functions are shifted from $J = 0$ to $J = J_{\rm c}$ with an increase in the current $J$. 
The vertical and horizontal axes are scaled by $\hbar n_{0}/g$ and $g n_{0}/\hbar$, respectively. 
The current $J$ is scaled by Landau's critical current $J_{0}$. 
} 
\label{Fig10.fig}
\end{figure}

We evaluate the local density spectral function in Bogoliubov theory for the uniform system. 
We consider a local density spectral function given by (\ref{EQ28}), 
where $n_{0} (x) = 1$ and $|G (x; {\bf k})|^{2}$ is also independent of $x$. 
In the low-energy regime, 
the local density spectral function is enhanced when $J$ increases (Figure~\ref{Fig10.fig}). 
The exponent of ${\mathcal I}_{n} (\omega)$ with respect to $\omega$ changes at $J = J_{\rm c}$. 

In the $d$-dimensional system for the stable superfluid state at $J < J_{\rm c}$, 
the low-energy dependence is given by 
\begin{align} 
{\mathcal I}_{n} (\omega) \simeq \frac{\Gamma_{d}}{2\pi} \frac{d  + J^{2}}{(1-J^{2})^{ (d+3)/2 }} \omega^{d}, 
\label{SMAblwqcBogo}
\end{align} 
where 
\begin{align}
( \Gamma_{1}, \Gamma_{2}, \Gamma_{3})
= 
\left ( 
1, \frac{1}{4}, \frac{1}{6 \pi}
\right ) . 
\label{GammadBogo}
\end{align} 
On the other hand, at $J = J_{\rm c} (=1)$, 
the density spectral function shows completely different behaviors. 
The low-energy behavior for the dimensionality $d$ is given by 
\begin{align}
{\mathcal I}_{n} (\omega) \simeq 
\frac{\Gamma_{d}'}{3 \pi} \omega^{(2d-3)/3}, 
\label{SMAeqqcBogo}
\end{align}
where 
\begin{align}
( \Gamma_{1}', \Gamma_{2}', \Gamma_{3}')
= 
\left ( 
1, \frac{2\sqrt{3}}{\pi}, \frac{1}{\pi}
\right ) . 
\label{GammadPrimeBogo}
\end{align} 
Derivations may be found in Appendix~\ref{Appendix7}. 
 
In a stable superfluid state $J < J_{\rm c} (= 1)$, the energy spectrum is a phonon, i.e., $E = (1 + J\cos\theta )k$. 
In the critical current state, 
$E \simeq k^{3}/8$ holds for low-$k$ when the momentum of the excitation is antiparallel to the supercurrent. 
The change of the energy spectrum from $E \propto k$ to $E \propto k^{3}$ 
increases the density of states, 
so that the density spectral function is enhanced at $J = J_{\rm c}$. 
This leads to the change of the exponent of the density spectral function with respect to $\omega$.

\section{Landau instability in Feynman's single-mode approximation }\label{SecVII}

Apart from mean-field theory, 
we reconsider the local density spectral function in the uniform system. 
We employ Feynman's single-mode approximation~\cite{Feynman1956}. 
We take $\hbar = 1$. 

The dynamic structure factor in Feynman's single-mode approximation is given by 
\begin{align}
S ({\bf q}, \omega)  = & \frac{q^2}{2 E_{\bf q}} \delta (\omega - E_{\bf q}) . 
\end{align} 
In fact, the relation between the energy of the elementary excitation $E_{\bf q}$ and the static structure factor $S({\bf q})$ 
is given by $E_{\bf q} = q^2/[2 S ({\bf q})]$, and we have a relation 
\begin{align}
\int_0^\infty d \omega S ({\bf q}, \omega) = \frac{q^{2}}{2 E_{\bf q}} . 
\end{align}

Even in the current flowing state, 
the strength of the dynamic structure factor is the same as that in the current free state 
because of translational invariance. 
In the current carrying state that flows along the $x$-direction, we end with 
\begin{align}
{\mathcal I}_{n} (\omega) 
= &\int \frac{d{\bf q}}{(2\pi)^d}
  \frac{q^2}{2 E_{\bf q}} \delta (\omega - E_{\bf q} - J q_{x}) , 
\end{align} 
where we used (\ref{eq11}) and 
\begin{align}
S ({\bf q}, \omega)  = & \frac{q^2}{2 E_{\bf q}} \delta (\omega - E_{\bf q} - J q_{x}). 
\end{align}  

For low $q = |{\bf q}|$, 
we suppose that 
\begin{align}
E_{\bf q} \simeq c_1 q + c_3 q^3 + {\mathcal O}(q^5) 
\label{Eq107New}
\end{align} 
holds, 
where $c_1$ and $c_3$ are positive coefficients. 
The analysis here focuses on the phonon regime, i.e., $c_1 q \gg c_3 q^3$. 
When $\omega_{-} \gg \omega$ (where $\omega_{-} = \sqrt{(c_{1} - J)^{3}/c_{3}}$), 
we have 
\begin{align}
{\mathcal I}_{n} (\omega) \simeq 
\frac{\Gamma_{d}}{2 \pi c_{1}} \frac{d c_{1}^{2} + J^{2}}{(c_{1}^{2} - J^{2})^{(d+3)/2}} \omega^{d}. 
\end{align} 
On the other hand, when $\omega_{-} \ll \omega \ll \omega_{+}$ (where $\omega_{+} = \sqrt{c_{1}^{3}/c_{3}}$), 
we obtain 
\begin{align}
{\mathcal I}_{n} (\omega) \simeq 
\frac{\Gamma_{d}'}{3\pi c_{1}} 
\frac{\omega^{(2d-3)/3}}{2^{(d+3)/2} J^{(d-1)/2} c_{3}^{(d+3)/6}}. 
\end{align} 
Details are provided in Appendix~\ref{Appendix7}. 
 
We finally discuss the local density spectral function for an ideal Bose gas, with the energy spectrum 
\begin{align}
E_{\bf k} = & \frac{ k^2}{2m}. 
\end{align} 
Let $|{\rm g} ; N \rangle$ be the $N$-particle ground state of the ideal Bose gas, 
where the $N$-particles occupy the single-particle ground state with ${\bf k} = 0$, 
and let $| l ; N\rangle$ be an excited state in the $N$-particle system. 
The matrix element is given by 
$\langle l; N | \hat n ({\bf r} = {\bf 0}) | {\rm g};N\rangle = \sqrt{N} / \Omega$, 
only when the excited state $l$ has momentum ${\bf k}$; 
otherwise, it becomes zero. 
Here, $\Omega$ is the system volume. 
This is because we have 
\begin{align}
\langle l; N | \hat n ({\bf r} = {\bf 0}) | {\rm g};N\rangle 
= & \langle l; N | \frac{1}{\Omega} \sum\limits_{{\bf k}, {\bf k}'} \hat a_{\bf k}^\dag \hat a_{\bf k '} | {\rm g};N\rangle 
\\
= & \frac{\sqrt{N}}{\Omega} \sum\limits_{{\bf k} } 
\langle l; N | \hat a_{\bf k}^\dag  | {\rm g};N-1\rangle, 
\end{align}
where 
$\hat a_{\bf k}$ is the annihilation operator of bosons and we used $
\hat a_{\bf k '} | {\rm g} ; N \rangle = \delta_{{\bf k'}, {\bf 0}} \sqrt{N} | {\rm g} ; N-1 \rangle 
$. As a result, 
the density spectral function of the ideal Bose gas 
is proportional to the density of states $D(\omega)$; that is, 
\begin{align}
{\mathcal I}_{n} (\omega) = \frac{N}{\Omega^{2}} D(\omega), 
\quad 
D(\omega) = \sum\limits_{\bf k} \delta ( \omega - E_{\bf k}).  
\end{align}  
We thus end with 
\begin{align}
{\mathcal I}_{n} (\omega) = \frac{N}{\Omega} \frac{ C_{d} m^{d} }{ 2^{(d+2)/2} \pi^{d}  } \omega^{(d-2)/2} 
\end{align}
in the $d$-dimensional system, where 
\begin{align}
(C_{1}, C_{2}, C_{3}) = (2, 2\pi, 4\pi). 
\end{align}

\section{Stability Criterion Hypothesis}\label{SecVIII}

We discuss the stability criterion hypothesis for superfluidity in light of the density spectral function ${\mathcal I}_{n}$~\cite{KatoWatabe2010JLTP,KatoWatabe2010PRL}, 
which is applicable to both the Landau instability and the instability of saddle-node bifurcation. 

We examined uniform systems in Sections~\ref{SecVI} and~\ref{SecVII}. 
The critical current $J_{\rm c}$ is equal to Landau's critical current.
For the stable superfluid ($J < J_{\rm c}$) in the system dimensionality $d$, 
${\mathcal I}_{n} \propto \omega^{d}$ holds. 
On the other hand, 
at $J = J_{\rm c}$, 
${\mathcal I}_{n} \propto \omega^{(2d -3)/3}$ holds, in which the exponent is less than the system dimensionality $d$. 
In the attractive external potential case discussed in Section~\ref{SecV}, 
the critical current is also equal to Landau's critical current. 
The low-$\omega$ behavior of ${\mathcal I}_{n}$ is the same as the results in this uniform system, 
although ${\mathcal I}_{n}$ involves an $x$-dependence. 

We also examined the local density spectral function 
in the presence of a repulsive potential wall in Section~\ref{SecV}. 
For a stable superfluid, the exponent of this function with respect to $\omega$ in the low-energy regime is equal to the system dimensionality $d$. 
On the other hand, 
for the critical current state, 
${\mathcal I}_{n} \propto \omega^{d-2}$ holds, in which the exponent is less than the system dimensionality $d$. 
Even if we calculate the density spectral function using an orthogonal basis instead of the tunneling solutions, 
these exponents will be unchanged as discussed in Section~\ref{SecV}.

In all cases discussed above, the exponent is equal to the system dimensionality for the stable superfluid state. 
For the critical current state, however, the exponent is less than the dimensionality,  
and this leads to the enhancement of the local density fluctuations in the low-energy regime. 
For the Landau instability, 
this enhancement originates from an anomaly in the energy spectrum, 
which leads to the enhancement of the density of states. 
For the soliton emission instability, 
the enhancement originates from an anomaly in the matrix element of the density fluctuations.  
All the results support the criterion~\cite{KatoWatabe2010JLTP,KatoWatabe2010PRL} 
\begin{eqnarray}
\lim_{\omega\rightarrow 0}{\mathcal I}_{n}({\bf r}, \omega) \propto 
\left \{
\begin{array}{ll}
\omega^{\beta}  & \qquad (J=J_{\rm c})
\\
\omega^{d} &\qquad (J<J_{\rm c}) 
\end{array}
\right. 
\end{eqnarray}
with $\beta < d$. 
The local density spectral function ${\mathcal I}_{n}({\bf r}, \omega)$ 
thus measures the vulnerability of superfluids.

We briefly discuss an ideal Bose gas. 
The ideal Bose gas is not a stable superfluid according to Landau's criterion. 
As examined in Section~\ref{SecVII}, 
the density spectral function of an ideal Bose gas is proportional to 
$\omega^{(d-2)/2}$. 
The exponent is less than the dimensionality $d$, 
so that the ideal Bose gas with $J = 0$ can be regarded as the critical current state 
according to our criterion. 
This is consistent with the Landau criterion.

The local density spectral function ${\mathcal I}_{n} ({\bf r}, t)$ 
is linked to the autocorrelation function $C_{n} ({\bf r}, t)$
according to (\ref{AutocorrlationCnrt}). 
An exponent of $\omega$ in the local density spectral function changes  in the low-energy regime at $J = J_{\rm c}$, 
An exponent of $t$ in the autocorrelation function also changes in the long-time regime. 
From the viewpoint of dimensional analysis, 
the autocorrelation function at large $t$ is given by 
\begin{eqnarray}
\lim_{t\rightarrow \infty} C_{n}({\bf r}, t) \propto 
\left \{
\begin{array}{ll}
1/t^{\beta+1} & \qquad  (J=J_{\rm c})
\\
1/t^{d+1}  &\qquad (J<J_{\rm c}) . 
\end{array}
\right. 
\label{Current5-39}
\end{eqnarray}

To demonstrate this behavior explicitly, we evaluate the autocorrelation function. 
We introduce the coarse-grained local density spectral function ${\mathcal I}_{n_{\rm CG}}({\bf r}, \omega)$ to eliminate unwanted high-frequency behavior. 
This function ${\mathcal I}_{n_{\rm CG}}({\bf r}, \omega) $ and the coarse-grained autocorrelation function $C_{n_{\rm CG}} ({\bf r}, t) $ are respectively 
given by 
\begin{align}
{\mathcal I}_{n_{\rm CG}}({\bf r}, \omega) 
= & 
\sum\limits_{l} 
| \langle l| \delta \hat{n}_{\rm CG} ({\bf r}) | {\rm g} \rangle |^{2} 
\delta (\omega - \omega_{l} + \omega_{\rm g}), 
\\
C_{n_{\rm CG}} ({\bf r}, t) 
= & 
\int d \omega {\mathcal I}_{n_{\rm CG}} ({\bf r}, \omega) \cos (\omega t)  . 
\end{align} 
Here, $\delta \hat n_{\rm CG} ({\bf r})$ is the coarse-grained local density fluctuation operator 
\begin{align}
\delta \hat n_{\rm CG} ({\bf r}) = 
\int d{\bf r}' f_{a} ({\bf r} - {\bf r}') \delta \hat n  ({\bf r}') , 
\end{align} 
where we take $\int d{\bf r}  f_{a} ({\bf r}) = 1$ and 
$f_{a} ({\bf r} ) \simeq 0$ for $|{\bf r}|\gg a$. 
One of the functions satisfying the above conditions is 
\begin{align}
f_{a} ({\bf r}) = \frac{1}{\pi^{d/2} a^{d}} \exp{( - |{\bf r}|^{2} / a^2 )}. 
\label{CGfa}
\end{align}

\begin{figure}[tbp]
\begin{center}
\includegraphics[width=6cm]{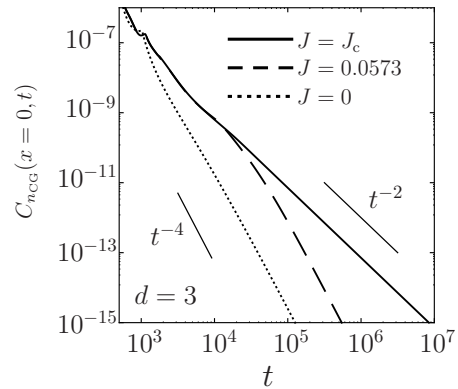}
\end{center}
\caption{
The coarse-grained autocorrelation function $C_{n_{\rm CG}} (x,t)$ 
at $x=0$ in the three-dimensional system with Bogoliubov theory. 
We employed the one-dimensional Gaussian potential barrier 
$V_{\rm ext} (x) = V_{0}\exp(-x^{2})$ with $V_{0} = 2$. 
The critical current in this case is $J_{\rm c} = 0.05740\cdots$. 
We used (\ref{CGfa}) with $a = 1$. 
The vertical and horizontal axes are scaled by $n_{0}^{2}$ and $\hbar / ( g n_{0} )$, respectively. 
$J$, $x$, $a$, and $V_{0}$ are scaled by $J_{0}$, $\xi$, $\xi$, and $gn_{0}$, respectively. 
} 
\label{Fig11.fig}
\end{figure}

The long-time behavior of the coarse-grained autocorrelation function 
for the critical current state is different than those for the other states at $J<J_{\rm c}$ (Figure~\ref{Fig11.fig}). 
The long-time behavior at $J (=J_{\rm c})$ is $t^{-2}$ and 
that at $J (<J_{\rm c})$ is $t^{-4}$. 
This is consistent with our criterion hypothesis (\ref{Current5-39}). 
In Figure~\ref{Fig11.fig}, we used the Gaussian-shaped potential barrier. 

We briefly comment on a related issue. 
In Tomonaga-Luttinger liquids, 
the autocorrelation function is given by~\cite{Giamarchi} 
\begin{align}
C_{n}({\bf r}, t) 
\sim \frac{A_{0}}{t^{2}} + \frac{A_{1}}{t^{2 K}}  + \frac{A_{2}}{t^{8 K}} + \cdots. 
\end{align}
$A_{0,1,2}$ are coefficients, and $K$ is the Tomonaga-Luttinger parameter. 
In the superconducting phase $(K>1)$, $C_{n}({\bf r}, t)  \propto  1/t^{2}$ holds for $t\rightarrow \infty$, 
and the exponent of $t^{-1}$ is $2$. 
On the other hand, in the charge-density wave (CDW) phase $(K<1)$, 
$C_{n}({\bf r}, t) \propto  1/t^{2K}$ holds for $t\rightarrow \infty$, 
and the exponent of $t^{-1}$ is $2K (< 2)$. 
In a one-dimensional system, 
the conductance in the superconducting phase is not infinity even when a small but non-zero voltage is applied~\cite{Giamarchi}, 
so that this does not completely correspond to the superfluidity discussed here. 
However, 
when we read the superconducting phase as the stable supercurrent state, 
and the CDW phase as the critical current state, 
the classification between the superconducting phase and the CDW phase is common to (\ref{Current5-39}).

\section{Conclusions}\label{SecVIIII}

A superflow through defects without dissipation is one of the most interesting superfluidity phenomena. 
Landau's criterion for superfluidity is developed by considering the elementary excitation energy 
on the basis of the Galilean transformation. 
Another mechanism of dissipation is the emissions of quantized vortices or solitons 
from an external potential.  
Through numerical calculations, 
these instabilities were categorized as a saddle node bifurcation. 
Thus, we aimed to understand the stability of superfluidity in both cases in an equal manner. 

In this paper, we studied the validity of the stability criterion hypothesis~\cite{KatoWatabe2010PRL,KatoWatabe2010JLTP}. 
This criterion states that the superfluid state is stable 
if an exponent of the local density spectral function ${\mathcal I}_{n}$ with respect to the energy (frequency) $\omega$ 
in the low-energy regime is equal to the system dimensionality $d$ (i.e., ${\mathcal I}_{n} \propto \omega^{d}$); 
however if it is less than $d$ (i.e., ${\mathcal I}_{n} \propto \omega^{\beta}$ with 
$\beta < d$), it is in the critical current state. 
This criterion indicates that the suppression of density fluctuations in the low-energy regime is a feature of a stable superfluid. 

Using Bogoliubov theory in the presence of a one-dimensional repulsive/attractive external potential, 
we evaluated the local density spectral function. 
Our numerical calculation using solutions of the tunneling problem and the orthogonal set 
supports the validity of the stability criterion hypothesis. 
Beyond Bogoliubov theory, 
we discussed the validity of this hypothesis in Feynman's single-mode approximation.

We can translate this criterion into autocorrelation function language. 
The criterion states that 
if the $t$-dependence of this function in the long-time regime is equal to $1/t^{d+1}$, 
then the superfluid state is stable. If it shows $1/t^{\beta+1}$ with $\beta < d$, it is in the critical current state. 
Evaluating the autocorrelation function in Bogoliubov theory, 
we numerically demonstrated this behavior 
in the presence of a one-dimensional repulsive potential wall.

We summarize interesting subjects for future studies. 
We have restricted ourselves to consider the system where the translational invariance holds in the $y$- and $z$-directions and these sizes are infinite. For the superfluid flowing in a capillary (or a channel), 
excitations at the surface are important for instabilities~\cite{Anglin2001,Fedichev2001}. 
Although numerical results for $d > 1$ demonstrated in this paper would not simply apply to such a realistic system, 
the enhancement of density fluctuations may appear at surface. 
We need to study the local density spectral function in the system with the transverse confinement (e.g. the system in Ref.~\cite{Engels2007}). 

Other prospective studies include confirming the criterion hypothesis for the vortex emission instability 
and applying the criterion to supersolidity in which translation invariance is broken. 
One may also ask whether the transport coefficients as well as other spectral and correlation functions 
(e.g. the current-current correlation function) show anomalous behavior in the critical current state. 
It would also be of interest to discuss the relation between the present criterion and 
the drag force~\cite{Sykes2009}, and to study autocorrelation functions in a strongly interacting Bose system beyond Bogoliubov theory 
in the presence of the potential barrier. 
 
\acknowledgements 
The authors thank D. Takahashi, Y. Nagai, M. Kunimi, T. Minoguchi, S. Sasa, M. Kobayashi, and H. Ohta 
for useful discussions. 
S. W. thanks G. Baym for discussion on the Landau instability, 
and also thanks J. Suzuki for discussions and comments. 
S.W. was supported by JSPS KAKENHI Grant Number (217751, 249416). 
This work was also supported by KAKENHI (21540352) from JSPS and KAKENHI (20029007) from MEXT in Japan.

\appendix 

\section{procedure to obtain tunneling solutions}\label{Appendix0}

We summarize the procedure for how to obtain the tunneling solutions of the Bogoliubov excitation. 
The tunneling of excitations in Bose--Einstein condensates through a potential barrier is referred to 
as the anomalous tunneling. 
These have been intensively and extensively studied 
for scalar Bose-Einstein condensates~\cite{Kovrizhin2001,Kagan2003,Danshita2006,Kato2007,Watabe2008,TsuchiyaOhahsi2008,Ohashi2008,Watabe2009RefleRefra,Takahashi2009II,TsuchiyaOhahsi2009,Takahashi2009,Takahashi2010}, 
and for Bose-Einstein condensates with internal degrees of freedom~\cite{WatabeKato2009,WatabeKato2011,WatabeKatoOhahsi2011A,WatabeKatoOhahsi2011B,WatabeKatoOhahsi2012}. 

We here consider the superfluid flowing through a one-dimensional potential barrier $V_{\rm ext}(x)$ with the current density $J$. 
The superflow is along the $x$-axis. 
The translational invariance holds in the $y$- and $z$-directions, and the potential barrier $V_{\rm ext}(x)$ is assumed to be localized around $x = 0$. 

In the tunneling problem, at the position far from the potential barrier, 
the wave function consists of the superposition of solutions in the homogeneous case. 
We first fix the incident energy $E$ as well as the incident angle $\theta$. 
This $\theta$ is an angle between the incident wave vector ${\bf k}^{\rm in} = (k_{x}^{\rm in}, k_{y}, k_{z})$ 
and the direction of the supercurrent density. 
After fixing $E$ and $\theta$, we determine the modulus of the incident wave vector $k^{\rm in} = |{\bf k}^{\rm in}|$ from a dispersion relation of the Bogoliubov excitation 
\begin{align}
E = k^{\rm in} J \cos{\theta} + \sqrt{\frac{(k^{\rm in})^{2}}{2} \left [ \frac{(k^{\rm in})^{2}}{2}  + 2 \right ]}. 
\end{align} 
Here, the modulus $k^{\rm in}$ is a positive and real solution of this equation. 
After the determination of $k^{\rm in}$, we fix $(k_{x}^{\rm in},k_{\perp}) = k^{\rm in} (\cos{\theta}, \sin{\theta})$, where $k_{\perp} \equiv \sqrt{k_{y}^{2} + k_{z}^{2}}$.

Once $E$ and $k_{\perp}$ are fixed, we can determine $k_{x}^{(1),(2),\pm}$ by solving (\ref{TB2-28}). 
$k_{x}^{(1)}$ is a real solution satisfying $k_{x}^{(1)} = k_{x}^{\rm in}$, and 
$k_{x}^{(2)}$ is the other real solution. 
The $k_{x}^{\pm}$ satisfy ${\rm sgn} ( {\rm Im} (k_{x}^{\pm}) ) = \pm 1$. 
We obtain a tunneling solution by solving the Bogoliubov equation (\ref{TB2-6}) with the boundary conditions (\ref{boundaryMP}) and (\ref{boundaryPM}). 
A practical approach to solving the Bogoliubov equation (\ref{TB2-6}) is to employ the finite element method~\cite{TsuchiyaOhahsi2009}. 
Indeed, we used this method to obtain the result of the autocorrelation function in Figure~\ref{Fig11.fig}, where a one-dimensional Gaussian-shaped potential barrier is employed. 
We have numerically calculated the local density spectral function in a one-dimensional Gaussian-shaped potential barrier case. 
The main results are the same as the $\delta$-function potential case as shown in this paper.

The one-dimensional $\delta$-function potential barrier case (i.e., $V_{\rm ext}(x) = V_{0} \delta (x)$) is the simplest case to determine the tunneling solution. 
We can use an analytic solution (\ref{eq47}). 
The wave function involving the incident and reflection waves is given by 
\begin{align}
{\bf u}_{\mp}(x) = & {\bf U}_{\mp}(x,k_{x}^{(1)}) + r {\bf U}_{\mp}(x,k_{x}^{(2)}) + a {\bf U}_{\mp}(x,k_{x}^{\mp}). 
\end{align}
The wave function involving the transmission wave is given by 
\begin{align}
{\bf u}_{\pm}(x) = & t {\bf U}_{\pm}(x,k_{x}^{(1)}) + b {\bf U}_{\pm}(x,k_{x}^{\pm}). 
\end{align}
Here, $r (t)$ is the amplitude reflection (transmission) coefficient, and ${\bf u}_{+}$ (${\bf u}_{-}$) is a solution at $x\geq 0$ $(x<0)$. 
The solution with the upper (lower) index is for the case at $0 \leq \theta < \pi/2$ ($\pi/2 < \theta \leq \pi$). 
We determine the coefficients $(r,t,a,b)$ from the boundary conditions (\ref{eq50}). 

We briefly note that when the real solutions $k_{x}^{(1)}$ and $k_{x}^{(2)}$ have the same sign, 
no reflection wave (i.e., double refraction) occurs. 
This condition can be reduced to 
\begin{align}
E < \sqrt{\frac{k_{\perp}^{2}}{2} \left ( \frac{k_{\perp}^{2}}{2} + 2 \right ) }, 
\label{Current5-27}
\end{align} 
because 
one of the relations between the solutions and coefficients with respect to (\ref{TB2-28}) 
is given by 
$k_{x}^{(1)} k_{x}^{(2)} |k_{x}^{+} |^{2} = k_{\perp}^{4} + 4 k_{\perp}^{2} - 4 E^{2}$, where we used $(k_{x}^{+})^{*} = k_{x}^{-}$. 
The region of $\theta$ satisfying (\ref{Current5-27}) is very narrow, and exists around $\theta = \pi/2$. 
In this case, 
we change the boundary conditions from (\ref{boundaryMP}) and (\ref{boundaryPM}) to 
\begin{align}
\begin{pmatrix}
u \\ v
\end{pmatrix} 
= & 
\tilde {\bf u}_{\mp} (x, k_{x}^{(1)}) 
+ 
a \tilde {\bf u}_{\mp} (x, k_{x}^{\mp})
\quad (x \rightarrow \mp \infty),  
\\
\begin{pmatrix}
u \\ v
\end{pmatrix}
=  & 
t  \tilde {\bf u}_{\pm} (x, k_{x}^{(1)}) 
+ 
r \tilde {\bf u}_{\pm} (x, k_{x}^{(2)}) 
+ 
b \tilde {\bf u}_{\pm} (x, k_{x}^{\pm}) 
\quad (x \rightarrow \pm \infty). 
\end{align} 
In the $\delta$-function potential case, at $0 \leq \theta < \pi/2$, 
we set 
\begin{align}
{\bf u}_{-}(x) = & {\bf U}_{-}(x,k_{x}^{(1)}) + a {\bf U}_{-}(x,k_{x}^{-}), 
\\
{\bf u}_{+}(x) = & t {\bf U}_{+}(x,k_{x}^{(1)}) + r {\bf U}_{+}(x,k_{x}^{(2)}) + b {\bf U}_{+}(x,k_{x}^{+}). 
\nonumber
\end{align}
At $\pi/2 < \theta \leq \pi$, we exchange the index $\pm$ in ${\bf u}$, ${\bf U}$, and $k_{x}$ for $\mp$.

\section{Wave functions of critical current state in the presence of an impurity potential}\label{Appendix2}

Here we derive the low-energy behavior of the function $G(x)$ in the critical current state. 
At the end of this appendix we obtain 
\begin{align}
\lim\limits_{E\rightarrow 0} 
G (x) = 
- \frac{2\sqrt{2} i}{\sqrt{k}}
C_{\rm III }^{(0)} A_{\varphi} (x) . 
\label{newAppendixA1}
\end{align} 
The technique for its derivation is based on~\cite{Takahashi2009,Takahashi2010}.

For the representation $(S,G)$, 
the equations in the presence of the one-dimensional potential barrier are given by 
\begin{align}
\hat{\mathcal H}_{\perp} S(x) 
-\frac{i J}{A(x)} \frac{d}{dx} \left [ \frac{G(x)}{A(x)} \right ] = & E G(x), 
\label{TB2-16} 
\\ 
\left [\hat{\mathcal H}_{\perp} + 2 A^{2} (x) \right ] G(x) 
- \frac{i J}{  A(x)} \frac{d}{dx} \left [ \frac{S(x)}{A(x)} \right ] = & E S(x), 
\label{TB2-17} 
\end{align} 
where $\hat{\mathcal H}_{\perp} = \hat{\mathcal H} + k_{\perp}^{2}/2$. 
Here, we used the translational invariance in the $y$- and $z$-directions, that is, 
\begin{align}
\begin{pmatrix}
S ({\bf r}) \\ G ({\bf r}) 
\end{pmatrix}
= 
\begin{pmatrix}
S (x) \\ G (x) 
\end{pmatrix}
e^{i(k_{y}y + k_{z}z)}. 
\end{align}

We are considering the supercurrent through a repulsive potential barrier, which corresponds to the condition $J_{\rm c} < 1$. 
In this case, the modulus of the incident momentum in the low-energy regime is linear in $E$, 
so that we have $k^{\rm in} = {\mathcal O} (E)$ as well as $k_{\perp} = {\mathcal O} (E)$. 
When we expand $S(x)$ and $G(x)$ with respect to the energy $E$, 
\begin{align}
S(x) = \sum\limits_{n=0}^{\infty} E^{n} S^{(n)}(x), \quad 
G(x) = \sum\limits_{n=0}^{\infty} E^{n} G^{(n)}(x), 
\label{Current5-2}
\end{align} 
we obtain equations for $n=0$: 
\begin{align}
\hat{\mathcal H}S^{(0)}(x) 
- \frac{iJ}{A(x)} \frac{d}{dx} \left [ \frac{G^{(0)}(x)}{A(x)} \right ] & = 0, 
\label{S0eqAppendix}
\\
\left [ \hat{\mathcal H} + 2A^{2}(x) \right ] G^{(0)}(x) 
- \frac{iJ}{A(x)} \frac{d}{dx} \left [ \frac{S^{(0)}(x)}{A(x)} \right ] & = 0, 
\label{G0eqAppendix}
\end{align} 
and those for $n = 1$: 
\begin{align}
\hat{\mathcal H}S^{(1)}(x) 
- \frac{iJ}{A(x)} \frac{d}{dx} \left [ \frac{G^{(1)}(x)}{A(x)} \right ] & = G^{(0)}(x), 
\label{S1eqAppendix}
\\
\left [ \hat{\mathcal H} + 2A^{2}(x) \right ] G^{(1)}(x) 
- \frac{iJ}{A(x)} \frac{d}{dx} \left [ \frac{S^{(1)}(x)}{A(x)} \right ] & = S^{(0)}(x). 
\label{G1eqAppendix}
\end{align} 
In this expansion (\ref{Current5-2}), we assumed that $S$ and $G$ start with ${\mathcal O} (E^{0})$, 
and omitted the normalization factor. 

We now consider the solutions $(S^{(0)}, G^{(0)})$. 
It is given by 
\begin{align}
\begin{pmatrix}
S^{(0)} (x) \\ G^{(0)} (x) 
\end{pmatrix} 
= 
\sum\limits_{j = {\rm I, II, III, IV}} C_{j} 
\begin{pmatrix}
S_{j} (x) \\ G_{j} (x) 
\end{pmatrix}, 
\label{newAppendixA7}
\end{align} 
where $C_{{\rm I, II, III, IV}} $ are coefficients, 
and $(S_{j} (x), G_{j} (x) )$ for $j = {\rm I, II, III, IV}$ are given by 
\begin{align}
\begin{pmatrix}
S_{\rm I}  \\ G_{\rm I} 
\end{pmatrix}
= & 
\begin{pmatrix}
A  \\ 0
\end{pmatrix} , 
\,\,
\begin{pmatrix}
S_{\rm II} \\ G_{\rm II} 
\end{pmatrix}
= 
\begin{pmatrix}
\displaystyle{ 
\hat P_{A} (1) - 2 i J \hat P_{A} \left ( G_{\rm II} / A \right )
}
\\
\displaystyle{ 
- 2 i J \hat P_{B} (A_{3})
}
\end{pmatrix} , 
\\
\begin{pmatrix}
S_{\rm III} \\ G_{\rm III} 
\end{pmatrix} 
= & 
\begin{pmatrix}
- 2 i q A A_{3}
\\
B 
\end{pmatrix} , 
\,\,
\begin{pmatrix}
S_{\rm IV}  \\ G_{\rm IV} 
\end{pmatrix} 
= 
\begin{pmatrix}
\displaystyle{ 
- 2 i J \hat P_{A} \left ( G_{\rm IV} / A \right ) 
}
\\
\displaystyle{ 
\hat P_{B} \left (1 \right )
}
\end{pmatrix}  . 
\end{align} 
Here, $A(x)$ is the amplitude of the condensate wave function determined by (\ref{TB2-9}), 
and $B(x)$ is an even parity solution of  
\begin{align}
\left [ \hat H + 2 A^{2} (x) - 2 \frac{J^{2}}{A^{4}(x)} \right ] B(x) =  0. 
\label{G0homoAppendix}
\end{align} 
We introduced 
\begin{align}
A_{3} (x) \equiv & \int_{0}^{x} dx ' \frac{B (x')}{A^{3} (x')}, 
\end{align} 
and 
\begin{align}
\hat P_{X} (Y) \equiv X(x) \int_{0}^{x} dx' \frac{Y(x')}{X^{2}(x')} . 
\end{align} 

Indeed, $S^{(0)}(x)$ and $G^{(0)}(x)$ are obtained as follows. 
The solution $S^{(0)}(x)$ is given by 
\begin{align}
S^{(0)}(x) = & C_{\rm I} S_{1} (x)+ C_{\rm II} S_{2} (x)+ f_{S} (x).  
\label{S0S1S2fS}
\end{align} 
Here, $S_{1}$ and $S_{2}$ are the general solutions of $\hat {\mathcal H} S^{(0)} (x) = 0$, given by 
\begin{align}
S_{1} (x) = A (x), \quad S_{2} (x) = A(x) \int_{0}^{x} \frac{dx'}{A^{2} (x')}. 
\end{align} 
A particular solution $f_{S}$ is 
\begin{align}
f_{S} (x) = - 2 i J A(x) \int_{0}^{x} dx' \frac{G^{(0)} (x')}{A^{3} (x')} . 
\end{align} 
where we used 
\begin{align}
f_{S} = & -2 \biggl ( 
-S_{1} \int dx' \frac{F_{S} S_{2}}{\Delta_{S}} + S_{2} \int dx' \frac{F_{S} S_{1}}{\Delta_{S}} 
\biggr ) 
\end{align}
with 
\begin{align}
F_{S}  = \frac{iJ}{A(x)} \frac{d}{dx} \left [ \frac{G^{(0)}(x)}{A(x)} \right ] 
\end{align} 
and the Wronskian $\Delta_{S} = S_{1} ( d S_{2} / dx) -  (d S_{1} / dx) S_{2} = 1$.

Substituting this result into (\ref{G0eqAppendix}), 
we obtain 
\begin{align}
\left [ \hat H + 2 A^{2} (x) - 2 \frac{J^{2}}{A^{4}(x)} \right ] G^{(0)} =  C_{\rm II} \frac{i J}{A^{3}(x)} \equiv F_{G} (x). 
\label{G0CIIAppendix}
\end{align} 
The solution $G^{(0)}(x)$ is given by 
\begin{align}
G^{(0)} (x) = & 
C_{\rm II} G_{\rm II} (x) + C_{\rm III} G_{1} (x) + C_{\rm IV} G_{2} (x) . 
\end{align}  
Here, $G_{1}$ and $G_{2}$ are the general solutions of (\ref{G0CIIAppendix}), given by 
\begin{align}
G_{1} (x) = B (x), \quad G_{2} (x) = B(x) \int_{0}^{x} \frac{dx'}{B^{2} (x')}. 
\end{align} 
A particular solution $f_{G}$ is 
\begin{align}
f_{G} (x) = - 2 i q C_{\rm II} B(x) \int_{0}^{x} dx' \frac{A_{3} (x')}{B^{2} (x')} , 
\end{align}
where we used 
\begin{align}
f_{G} = & 
-2 \biggl ( 
-G_{1} \int dx' \frac{F_{G}G_{2}}{\Delta_{G}} + G_{2} \int dx' \frac{F_{G}G_{1}}{\Delta_{G}} 
\biggr ) 
\end{align} 
and the Wronskian $\Delta_{G} = 1$. 
For $G_{\rm II}$, we defined $G_{\rm II} \equiv f_{G} /C_{\rm II}$. 
Substituting this solution $G^{(0)} (x)$ into (\ref{S0S1S2fS}), 
we obtain (\ref{newAppendixA7}).

At $J = J_{\rm c}$, we find 
\begin{align}
B(x) = & \frac{\partial A(x)}{\partial \varphi} \equiv A_{\varphi} (x). 
\end{align} 
In fact, (\ref{G0homoAppendix}) has the same form as 
\begin{align}
\left  [ \hat{\mathcal H} + 2 A^{2}(x) - \frac{2J_{\rm c}^{2}}{A^{4}(x)} \right ] 
\frac{\partial A(x) }{ \partial \varphi}
= & J_{\rm c} \frac{dJ}{d\varphi} \left [ A(x) - \frac{1}{A^{3}(x)} \right ] 
\nonumber
\\
= & 0. 
\label{Current5-1}
\end{align} 
This equation is obtained from (\ref{TB2-9}), 
where we took the derivative with respect to $\varphi$. 
We also used the relation $\partial J/\partial \varphi = 0$ which is correct only at $J = J_{\rm c}$. 
Since 
\begin{align}
\left ( - \frac{1}{2} \frac{d^{2}}{dx^{2}} + 2 - 2 J^{2} \right ) B(x) = 0, 
\end{align}
at $|x| \gg 1$, 
$B (x)$ is given by 
\begin{align}
B (x) = \beta e^{- \kappa |x|}, 
\end{align} 
at $|x| \gg 1$ with $\beta$ being a constant and $\kappa = 2 \sqrt{1 - J^{2}}$. 

Note that 
$(S_{\rm I,III} , G_{\rm I,III})$ converge, but $(S_{\rm II,IV} , G_{\rm II,IV})$ exponentially diverge at $|x| \rightarrow \infty$. 
Indeed, we obtain 
\begin{align}
(S_{\rm II} , G_{\rm II} )  
\simeq  & 
\left (- 2 \frac{J^{2}\alpha_{3}}{\beta \kappa^{2}} e^{\kappa |x|} {\rm sgn} (x) , - \frac{iJ\alpha_{3} }{\beta \kappa } e^{\kappa |x|} \right  ) , 
\\
(S_{\rm IV} , G_{\rm IV}) 
\simeq &  
\left ( 
- \frac{iJ}{\kappa^{2} \beta} e^{\kappa |x|} , 
\frac{e^{\kappa |x|}}{2 \kappa \beta} {\rm sgn} (x) 
\right ), 
\end{align}
where 
\begin{align}
\alpha_{1} \equiv  A_{1} (\infty), \quad \alpha_{3} \equiv  A_{3} (\infty) , \quad 
\eta \equiv \alpha_{1} / \alpha_{3} , 
\end{align} 
with 
\begin{align}
A_{1} (x) =  & \int_{0}^{x} dx' A(x') B (x') . 
\end{align}  
At $|x| \gg 1$, we have $A_{j} = {\rm sgn} (x) [ \alpha_{j} - (\beta / \kappa) e^{-\kappa |x|} ]$ for $j = 1$ and $3$.

However, as shown below,  
the particular solutions for $n = 1$, generally given by 
\begin{align}
S_{\rm p}^{(1)} (x) = & 
- 2 \hat P_{A} \biggl ( 
\int_{0}^{x} dx' A(x')  G^{(0)} (x')
\biggr )
\nonumber 
\\ & 
- 2 i J \hat P_{A} ( G_{\rm p}^{(1)} / A) , 
\label{particularS}
\\
G_{\rm p}^{(1)} (x) = & - 2 \hat {P}_{B} 
\biggl ( 
\int_{0}^{x} dx' B(x') \biggl [ S^{(0)} (x') 
\nonumber 
\\ & 
- \frac{2 i q}{A^{3}(x')} \int_{0}^{x'} dx'' A(x'') G^{(0)} (x'') \biggr ]
\biggr ) ,  
\label{particularG}
\end{align} 
cancel out the divergences in $(S_{\rm II,IV} , G_{\rm II,IV})$. 

We first consider a set of particular solutions $(G_{\rm p, I}^{(1)}, S_{\rm p, I}^{(1)})$ 
where $(S^{(0)}, G^{(0)})$ is given by $(S_{\rm I}, G_{\rm I})$. 
At $|x|\gg 1$, 
we have 
\begin{align}
(S_{\rm p, I}^{(1)}, G_{\rm p, I}^{(1)}) 
\simeq 
\left ( 
- \frac{2 iJ  \alpha_{1}}{\kappa^{2} \beta} {\rm sgn}(x) e^{\kappa |x|} 
, 
- \frac{\alpha_{1}}{ \kappa \beta} e^{\kappa |x|}
\right ), 
\end{align}
where we used 
\begin{align}
G_{\rm p, I}^{(1)} (x) = & - 2 B(x) \int_{0}^{x} d x' \frac{A_{1} (x')}{B^{2}(x')} , 
\\
S_{\rm p, I}^{(1)} (x) = & - 2 i J A(x) \int_{0}^{x} dx' \frac{G_{\rm p}^{(1)}(x')}{A^{3} (x')} . 
\end{align}  

In the case where $(S^{(0)}, G^{(0)})$ is given by $(S_{\rm III}, G_{\rm III})$, 
a set of particular solutions $( G_{\rm p, III}^{(1)} , S_{\rm p, III}^{(1)} )$ at $|x|\gg 1$ 
is given by 
\begin{align}
( S_{\rm p,III}^{(1)}, G_{\rm p,III}^{(1)} ) 
\simeq 
\left ( 
 \frac{4 J^{2}  \alpha_{1} \alpha_{3}}{\beta^{ }  \kappa^{2} } e^{  \kappa |x|} , 
 \frac{2i J  \alpha_{1} \alpha_{3}}{\beta^{} \kappa } e^{  \kappa |x|}  {\rm sgn}(x) 
\right ) , 
\end{align}
where we used 
\begin{align}
 G_{\rm p,III}^{(1)} (x) 
= & 
4 i J \hat {P}_{B} ( A_{1} A_{3} ) ,
\\ 
S_{\rm p ,III}^{(1)}  (x)  = & -2 \hat {P}_{A} (A_{1} )  
- 2 i J \hat {P}_{A} ( G_{\rm p, III}^{(1)} / A) . 
\end{align}  

As a result, from the combination of $(S_{\rm II, IV}, G_{\rm II, IV})$ and $(S_{\rm p,I,III}^{(1)}, G_{\rm p,I,III}^{(1)})$, 
we can construct solutions without exponential divergences, given by 
\begin{align} 
\begin{pmatrix} S_{\rm I}^{(1)} (x) \\ G_{\rm I}^{(1)} (x) \end{pmatrix}
= & 
\begin{pmatrix} S_{\rm p,I}^{(1)} (x) \\ G_{\rm p,I}^{(1)} (x) \end{pmatrix}
- \frac{\eta}{i J}
\begin{pmatrix} S_{\rm II}^{} (x) \\ G_{\rm II}^{} (x) \end{pmatrix} , 
\\
\begin{pmatrix} S_{\rm III}^{(1)} (x) \\ G_{\rm III}^{(1)} (x) \end{pmatrix}
= & 
\begin{pmatrix} S_{\rm p,III}^{(1)} (x) \\ G_{\rm p,III}^{(1)} (x) \end{pmatrix}
- 4 i J \alpha_{1} \alpha_{3}
\begin{pmatrix} S_{\rm IV}^{} (x) \\ G_{\rm IV}^{} (x) \end{pmatrix} . 
\end{align} 
Indeed, $(S_{\rm I,III}^{(1)}, G_{\rm I,III}^{(1)} )$ at $|x| \gg 1$ are given by 
\begin{align}
S_{\rm I}^{(1)} \simeq & - \frac{\eta}{iJ} [ x + \gamma {\rm sgn}(x)]  - \frac{iJ (1-\eta)}{1 - J^{2}} [ x + \nu {\rm sgn}(x)] , 
\nonumber 
\\
G_{\rm I}^{(1)} \simeq & \frac{1 - \eta}{2 (1-J^{2})} , 
\nonumber 
\\
G_{\rm III}^{(1)}  
\simeq & - \alpha_{3} \frac{iJ (1 + \eta)}{1-J^{2}} {\rm sgn} (x) , 
\nonumber 
\\
S_{\rm III}^{(1)} 
\simeq & 
\alpha_{3} \left [ 
- 2 \eta (|x| + \lambda) - 2 \frac{J^{2}(1+\eta)}{1-J^{2}} (|x| + \nu)
\right ] . 
\nonumber 
\end{align}
Here, $\lambda$ is a constant, 
and $\gamma$ and $\nu$ are respectively given by 
$\gamma \equiv A (x) \int_{0}^{\infty} dx' \left [ A^{-2} (x') - 1 \right ]$ 
and 
$\nu \equiv \int_{0}^{\infty} dx' \left [ A^{-3} (x') - 1 \right ]$. 

As a result, the solutions with the first order of $E$ without exponential divergences 
are given by 
\begin{align}
\begin{pmatrix}
S_{\rm I,III}^{\rm total} \\ G_{\rm I,III}^{\rm total}
\end{pmatrix} 
= & 
\begin{pmatrix}
S_{\rm I,III}^{(0)} \\ G_{\rm I,III}^{(0)} 
\end{pmatrix} 
+ 
E 
\begin{pmatrix}
S_{\rm I,III}^{(1)} \\ G_{\rm I,III}^{(1)}
\end{pmatrix} 
+ {\mathcal O}(E^{2}) . 
\end{align}
In particular, $S_{\rm I, III}^{\rm total}$ behave as 
\begin{align}
S_{\rm I}^{\rm total} = & 1 + E \left [ \frac{J^{2} - \eta}{iJ (1-J^{2})} x + \tilde \gamma {\rm sgn}(x) \right ], 
\\
\frac{ S_{\rm III}^{\rm total} }{- 2 i J \alpha_{3}}
= & 
{\rm sgn} (x) + \frac{J^{2}+\eta}{iJ (1 - J^{2})} E |x| + \tilde \lambda E, 
\end{align}
with 
\begin{align}
\tilde \gamma = & 
- \frac{1}{iJ} \left [ \eta \gamma - \frac{J^{2} (1-\eta)}{1-J^{2}}\nu \right ], 
\\
\tilde \lambda = & 
\frac{1}{i J} \left [  \eta \lambda + \frac{J^{2} (1+\eta)}{1-J^{2}} \kappa \right ] . 
\end{align} 

We replace $C_{\rm III}$ by $C_{\rm III}/(- 2 i J \alpha_{3})$. 
In fact, $C_{\rm III}$ is just a coefficient to be determined later. 
In this case, 
we end with 
\begin{align}
\begin{pmatrix}
S (x) \\ G (x) 
\end{pmatrix} 
= 
C_{\rm I}
\begin{pmatrix}
S_{\rm I}^{\rm total} (x) \\ G_{\rm I}^{\rm total} (x) 
\end{pmatrix} 
-  
\frac{C_{\rm III}}{ 2 i J \alpha_{3}} 
\begin{pmatrix}
S_{\rm III}^{\rm total} (x) \\ G_{\rm III}^{\rm total} (x) 
\end{pmatrix} . 
\label{AppendixSSISIII}
\end{align} 

$A_{3}$ at $J_{\rm c}$ and the phase difference $\varphi$ are given by 
\begin{align}
A_{3} (x) = & \int_{0}^{x} dx' \frac{A_{\varphi}(x')}{A^{3} (x')} 
= - \frac{1}{2} \frac{\partial }{\partial \varphi} \int_{0}^{x} dx' \left [ \frac{1}{A^{2} (x')} - 1 \right ], 
\\
\varphi = & \frac{J}{2} \int_{0}^{\infty} dx' \left [ \frac{1}{A^{2} (x')} - 1 \right ]. 
\end{align} 
We then obtain 
\begin{align}
\alpha_{3} = A_{3} (\infty) = - \frac{1}{2} \frac{\partial }{\partial \varphi} \left ( \frac{\varphi}{2J} \right ) = - \frac{1}{4J} . 
\end{align}
As a result, the factor $- 2 i J \alpha_{3}$ can be reduced into $- 2 i J \alpha_{3} = i/2$. 
$\eta = {\mathcal O} (J)$ also holds. 

At $|x| \gg 1$, the low energy behavior of $S$ is 
\begin{align}
S = &C_{\rm I}^{(0)} + C_{\rm III}^{(0)} {\rm sgn} (x) 
\nonumber
\\ & + E 
[
C_{\rm I}^{(1)} + C_{\rm I}^{(0)} \tilde\gamma {\rm sgn} (x) + C_{\rm III}^{(1)} {\rm sgn} (x) 
+ \tilde\lambda C_{\rm III}^{(0)}
]
\nonumber
\\ & + E x 
[
C_{\rm I}^{(0)} \frac{J^{2} - \eta}{iJ (1- J^{2})} + C_{\rm III}^{(0)} \frac{J^{2}+\eta}{i J (1-J^{2})} {\rm sgn} (x)
] . 
\label{SLowEnergyAppendix}
\end{align} 
We here expanded $C_{\rm I, III}$ by energy $E$, i.e., 
$C_{\rm I,III} = C_{\rm I,III}^{(0)} + E C_{\rm I,III}^{(1)} + {\mathcal O} (E^{2})$. 
This form will be used to determine the coefficients in the tunneling problem, 
which will be examined in Appendix~\ref{Appendix4}. 

So far, we have assumed that the wave function in the low-energy regime 
starts with ${\mathcal O} (E^{0})$. 
However, $S$ and $G$ in the uniform system are given by 
\begin{align}
\begin{pmatrix}
S (x)  \\ G (x)
\end{pmatrix}
= 
\begin{pmatrix}
\alpha_{ k_{x} }\\ \beta_{ k_{x} }
\end{pmatrix}  
e^{ik_{x}x}
, \quad 
\begin{pmatrix}
\alpha_{ k_{x} }\\ \beta_{ k_{x} }
\end{pmatrix} 
= 
\frac{1}{\sqrt{ {\rm Re}[{\cal M}] }}
\begin{pmatrix}
1 \\ {\cal M}
\end{pmatrix}, 
\label{NormalizedSG}
\end{align}
where the normalization coefficient ${\cal M}$ is 
\begin{align}
{\cal M} = \frac{ k_{x}^{2} + k_{\perp}^{2} }{2 (E - J k_{x} )}, 
\end{align} 
so that (\ref{NormalizedSG}) satisfies $(SG^{*} + S^{*}G)/2 =1$. 
In the low-momentum regime, $S \simeq \sqrt{2/k}$ holds where $k = \sqrt{k_{x}^{2} + k_{\perp}^{2}}$. 
Although $G \propto \sqrt{k/2} $ holds, this is true only for the uniform system. 
According to (\ref{AppendixSSISIII}), $G(x)$ in the critical current state starts with the same order as 
$S(x)$ with respect to $E$. 
As a result, when we calculate physical quantities, such as the density spectral function, we should multiply 
(\ref{AppendixSSISIII}) by the factor $\sqrt{2/k}$. 
At the critical current, 
$\lim_{E\rightarrow 0} G_{\rm I}^{\rm total} (x) = 0$ 
and $\lim_{E\rightarrow 0} G_{\rm III}^{\rm total} (x) = B(x) = A_{\varphi} (x)$ hold.  
We then end with 
\begin{align}
\lim\limits_{E\rightarrow 0} 
G (x) = & \sqrt{\frac{2}{k}} \frac{C_{\rm III}^{(0)} }{- 2 i J \alpha_{3}} A_{\varphi} (x). 
\end{align} 
This leads to (\ref{newAppendixA1}).

\section{Local density spectral function in the critical current state for soliton instability}\label{Appendix4}

We evaluate the local density spectral function 
in the critical current state in the presence of a repulsive potential barrier at the low-energy limit. 
The goal in this appendix is to derive (\ref{Current5-33}). 

We start with the case of a system dimensionality $d = 1$. 
When the incident excitation is the right (left)-moving one, 
we find 
\begin{align}
k_{x}^{(1)} = \pm \frac{E}{1 \pm J}, \quad k_{x}^{(2)} = \mp \frac{E}{1 \mp J}. 
\end{align} 
The boundary condition at $|x| \gg 1$ with incident and reflection waves 
and that with a transmission wave 
can be reduced to 
\begin{align}
S (x) = & \exp{[ ik_{x}^{(1)}x ]} + r \exp{[ ik_{x}^{(2)}x ]} 
\\
\simeq &  1 + r^{(0)} + E r^{(1)} 
+ E x \left [ \frac{i}{\pm 1+J} + \frac{ir^{(0)}}{\mp 1 + J} \right ] , 
\nonumber 
\\
S (x) = & t \exp{[ ik_{x}^{(1)} x ]} 
\simeq 
t^{(0)} + E t^{(1)} 
+ E x  \frac{it^{(0)}}{\pm 1 + J} . 
\end{align} 
Here, we expanded coefficients as $t \simeq t^{(0)} + E t^{(1)} + {\mathcal O}(E^{2})$ and 
$r \simeq r^{(0)} + E r^{(1)} + {\mathcal O}(E^{2})$. 
Comparing coefficients in (\ref{SLowEnergyAppendix}) with those in the above equations, 
we end with 
\begin{align}
\begin{pmatrix}
t^{(0)} \\ r^{(0)}
\end{pmatrix}
= 
\begin{pmatrix}
\displaystyle{ \frac{ \mp 2 J \eta}{\eta^{2} + J^{2}} } 
\\ 
\displaystyle{ \frac{J^{2} - \eta^{2}}{\eta^{2} + J^{2}} }
\end{pmatrix}, 
\quad 
\begin{pmatrix}
C_{\rm I}^{(0)}  \\ C_{\rm III}^{(0)} 
\end{pmatrix}
= 
\begin{pmatrix}
\displaystyle{ \frac{J (J \pm \eta)}{J^{2} + \eta^{2}}  } 
\\ 
\displaystyle{ \mp \frac{J (J \mp \eta)}{J^{2} + \eta^{2}} }
\end{pmatrix} . 
\nonumber
\end{align} 
The coefficients in the case of the right-moving incident excitation $C_{\rm III, R}^{(0)}$ and the left-moving incident excitation $C_{\rm III, L}^{(0)}$ can be summarized as 
\begin{align}
C_{\rm III, R}^{(0)} = - \frac{J (J - \eta) }{J^{2} + \eta^{2}}, 
\quad 
C_{\rm III, L}^{(0)} = \frac{J (J + \eta) }{J^{2} + \eta^{2}} . 
\end{align}
As a result, the local density spectral function in the one-dimensional system is given by 
\begin{align}
{\mathcal I}_{n} (\omega) = & 
\frac{n_{0} (x)}{2\pi} \int_{-\infty}^{0} d k_{x} 
\left | \sqrt{\frac{2}{|k_{x}|}} \frac{C_{\rm III, L }^{(0)} }{- 2 i J \alpha_{3}} A_{\varphi} (x) \right |^{2}
\nonumber 
\\ & \times 
\delta (\omega - |k_{x}| - J k_{x})
\nonumber 
\\ & 
+ 
\frac{n_{0} (x)}{2\pi} \int_{0}^{\infty} d k_{x} 
\left |\sqrt{\frac{2}{|k_{x}|}} \frac{C_{\rm III, R }^{(0)} }{- 2 i J \alpha_{3}} A_{\varphi} (x) \right |^{2}
\nonumber 
\\ & \times 
\delta (\omega - |k_{x}| - J k_{x})
\\ 
= & 
\frac{2J^{2}}{\pi (J^{2} + \eta^{2})} \frac{1}{\omega} [ \partial_{\varphi} n_{0} (x) ]^{2} . 
\end{align} 
This is just (\ref{Current5-33}) for a dimensionality $d = 1$.

Next, we consider the two- and three-dimensional systems. 
In the low-energy regime, 
the energy spectrum is given by 
\begin{align}
E \simeq J k^{\rm in} \cos \theta^{} + k^{\rm in}. 
\end{align}
As a result, we obtain 
\begin{align}
k^{\rm in} = \frac{E}{1 + J \cos \theta_{ }}, \quad 
k_{x}^{\rm in} = \frac{E \cos \theta_{ }}{1 + J \cos \theta_{}} , 
\end{align}
and $k_{\perp} = k^{\rm in} \sin \theta_{}$. 
In the low-energy regime, we also have 
\begin{align}
E = J k_{x} + \sqrt{k_{x}^{2} + k_{\perp}^{2}}. 
\end{align} 
Solving this equation with respect to $k_{x}$, we obtain $k_{x} \simeq \pm E \cos \theta $. 
Here, we assumed that the potential barrier is strong, which leads to $J = J_{\rm c} \ll 1$. 
We also considered the low-energy regime, so that we can take $J E \ll 1$. 

The incident and reflection momenta $k_{x}^{(1)}$ and $k_{x}^{(2)}$ are now given by 
\begin{align}
k_{x}^{(1)}  = + E \cos \theta, 
\quad 
k_{x}^{(2)}  = - E \cos \theta. 
\end{align} 
In the low-energy regime, 
the boundary condition at $|x| \gg 1$ with incident and reflection waves 
and that with a transmission wave 
can be reduced to 
\begin{align}
S (x) 
\simeq & 1 + r^{(0)} + E r^{(1)}  
+ Ex i [1 - r^{(0)}] \cos \theta, 
\\
S (x) 
\simeq & t^{(0)} + E t^{(1)} + Ex i t^{(0)} \cos \theta. 
\end{align} 

Comparing coefficients in (\ref{SLowEnergyAppendix}) with those in the above equations, 
we find  
\begin{align}
\begin{pmatrix}
t^{(0)}  \\ r^{(0)} 
\end{pmatrix}
= & 
\begin{pmatrix}
\displaystyle{ 
\frac{ 2 \eta J \cos \theta_{} } {J^{2} \cos^{2} \theta_{ } + \eta^{2} }
}
\\ 
\displaystyle{ 
\frac{J^{2} \cos^{2} \theta_{ } - \eta^{2}} {J^{2} \cos^{2} \theta_{ } + \eta^{2} }
}
\end{pmatrix}, 
\\
\begin{pmatrix}
C_{\rm I}^{(0)}  \\ C_{\rm III}^{(0)}
\end{pmatrix}
= & 
\begin{pmatrix}
\displaystyle{ 
\frac{J \cos \theta_{ } (\eta + J \cos \theta_{ })} {J^{2} \cos^{2} \theta_{ } + \eta^{2} }
}
\\ 
\displaystyle{ 
\pm \frac{J \cos \theta_{ } (\eta - J \cos \theta_{ })} {J^{2} \cos^{2} \theta_{ } + \eta^{2} }
}
\end{pmatrix} , 
\end{align}
where the upper sign is for $ 0 \leq  \theta < \pi/2$ and the lower sign is for $ \pi/2 <  \theta  \leq \pi$.  

In the two- and three-dimensional systems for a low-energy regime, 
${\mathcal I}_{n} (x,\omega)$ can be reduced to 
\begin{align}
{\mathcal I}_{n} (x,\omega) \simeq 2 [ \partial_{\varphi} n_{0} (x) ]^{2} 
W (\omega), 
\label{newAppendixC34}
\end{align}
where 
\begin{align}
W (\omega) =
\int \frac{d {\bf k}}{(2\pi)^{d}} 
\frac{1}{k} |C_{\rm III}^{(0)}|^{2} 
 \delta (\omega - k - kJ \cos\theta) . 
\end{align} 
In the two-dimensional system, 
we have 
\begin{align}
W (\omega) 
\simeq & 2 \int_{0}^{\infty} \frac{dk}{(2\pi)^{2}} k \int_{0}^{\pi} d \theta 
\frac{1}{k} |C_{\rm III}^{(0)}|^{2} 
\nonumber 
\\ & \times 
\frac{1}{1+J\cos \theta_{\rm in}} \delta \left ( k - \frac{\omega}{1 + J \cos \theta_{\rm in}} \right )
\\
\simeq & 
\frac{1}{2\pi} \left ( 1 - \frac{\eta}{\sqrt{J^{2} + \eta^{2}}} \right ). 
\end{align} 
In the three-dimensional system, 
we have 
\begin{align}
W (\omega) 
\simeq & 2 \pi \int_{0}^{\infty} \frac{dk}{(2\pi)^{3}} k^{2} \int_{0}^{\pi} d \theta \sin \theta
\frac{1}{k} |C_{\rm III}^{(0)}|^{2} 
\nonumber 
\\ & \times 
\frac{1}{1+J\cos \theta_{ }} \delta \left ( k - \frac{\omega}{1 + J \cos \theta_{ }} \right )
\\ 
\simeq & 
\frac{\omega}{2 \pi^{2}} 
\left [ 
1 - \frac{\eta}{J} \tan^{-1} \left ( \frac{J}{\eta} \right ) 
\right ] . 
\end{align} 
In conjunction with (\ref{newAppendixC34}), 
we obtain (\ref{Current5-33}) for the dimensionalities $d = 2$ and $3$.

\section{Spectral functions in Feynman's single-mode approximation}\label{Appendix7}

We evaluate the local density spectral function in a $d$-dimensional system 
within Feynman's single-mode approximation 
\begin{align}
{\mathcal I}_{n} (\omega) = & 
\int \frac{d{\bf q}}{(2\pi)^d} 
\frac{q^2}{2 E_{\bf q}} \delta (\omega - E_{\bf q} - J q_{x}). 
\end{align} 
At the end of this appendix, 
we will discuss the local density spectral function within Bogoliubov theory in a uniform system for dimensionality $d$.

We first evaluate the one-dimensional system, where the spectral function is given by 
\begin{align}
{\mathcal I}_{n} (\omega) = & \frac{1}{2\pi} 
\int_{-\infty}^\infty d q_x \frac{q_x^2}{ 2E_{q_x} } 
\delta (\omega - E_{q_{x}} - J q_x). 
\end{align} 
Let $k_\pm (\omega)$ be solutions of 
\begin{align}
\omega = E_{\bf q} \pm J q = f_\pm (q) . 
\end{align}
In this case, we obtain 
\begin{align}
{\mathcal I}_{n} (\omega) = &
\frac{1}{2\pi }  \int_{-\infty}^\infty d q_x  
\sum\limits_{j = \pm}
\frac{q_x^2}{ 2E_{q_x} } 
 \delta (q_x - j k_j ) \left | \frac{\partial f_j}{\partial q} \right |_{q = k_j }^{-1} 
\nonumber 
\\
= & 
\frac{1}{4\pi} \sum\limits_{j = \pm} \frac{k_j^2 (\omega)}{E_{k_j (\omega)} }  
\frac{d k_j (\omega) }{d \omega} , 
\end{align}
where we used 
\begin{align}
 \left | \frac{\partial f_j}{\partial q} \right |_{q = k_j (\omega) }^{-1} = 
 \frac{d k_j (\omega) }{d \omega} . 
\end{align}

We suppose that the energy spectrum $E_{\bf q}$ is given by (\ref{Eq107New}) for low $q = |{\bf q}|$, 
and the low-energy excitation is a phonon, i.e., $c_1 q \gg c_3 q^3$. 
In this case, we obtain 
\begin{align}
E_{\bf q} = c_{1} q, 
\quad 
k_{+} (\omega) \simeq \frac{\omega}{c_{1} + J}. 
\end{align} 
We thus end up with 
\begin{align}
{\mathcal I}_{n} (\omega) = & 
\frac{1}{8 \pi c_1} \sum\limits_{j = \pm}  \frac{d k_j^2 (\omega) }{d \omega} . 
\end{align}

When $q_{-} \ll q$ with $q_{-} = \sqrt{ (c_{1} - J)/ c_{3}}$, 
we obtain 
\begin{align}
k_{-} (\omega) \simeq \left ( \omega / c_{3} \right )^{1/3}. 
\end{align} 
The condition $q_{-} \ll q$ can be reduced to 
$\omega_{-} \ll \omega$ with $\omega_{-}= \sqrt{ (c_{1} - J)^{3}/ c_{3}}$. 
On the other hand, when $q_{-} \gg q$, 
we obtain 
\begin{align}
k_{-} (\omega) \simeq \omega / (c_{1} - J) . 
\end{align}
The condition $q_{-} \gg q$ can be reduced to $\omega_{-} \gg \omega$.  

As a result, when $\omega_{-} \gg \omega$, 
we obtain 
\begin{align}
{\mathcal I} (\omega )  = & \frac{1}{8\pi c_{1}} \frac{d}{d\omega} \left [ \left ( \frac{\omega}{c_{1} - J} \right )^{2} + \left ( \frac{\omega}{c_{1} + J} \right )^{2} \right ] 
\\
= & \frac{\omega}{2\pi c_{1}} \frac{ (c_{1}^{2} + J^{2})}{(c_{1}^{2} - J^{2})^{2}} . 
\label{newAppendixF13}
\end{align}
On the other hand, when $\omega_{-} \ll \omega \ll \omega_{+}$ with 
$\omega_{+}= \sqrt{ c_{1}^{3}/ c_{3}}$, 
we obtain 
\begin{align}
{\mathcal I} (\omega )  = & \frac{1}{8\pi c_{1}} \frac{d}{d\omega} \left [ \left ( \frac{\omega}{c_{3}} \right )^{2/3} + \left ( \frac{\omega}{c_{1} + J} \right )^{2} \right ] 
\\
\simeq & \frac{1}{12\pi c_{1}} \frac{\omega^{-1/3}}{ c_{3}^{2/3}} . 
\label{newAppendixF15}
\end{align}

We evaluate the spectral function in the two-dimensional system where 
\begin{align}
{\mathcal I}_{n} (\omega) 
= & \frac{1}{(2\pi)^{2}} 
\int d q q \int_{0}^{2\pi} d  \theta
\frac{q^2}{ 2E_{q} } 
\delta (\omega - E_{q} - J q \cos \theta) . 
\nonumber 
\end{align}
The condition where the equation in the delta-function is zero is given by 
$\left | ( \omega - E_{q} ) / (Jq) \right | \leq 1$. 
This condition can be reduced to 
$k_{+} (\omega ) \leq q \leq k_{-} (\omega)$.  
Then, we obtain 
\begin{align}
{\mathcal I}_{n} (\omega) 
= & \frac{2}{(2\pi)^{2}} 
\int_{k_+ (\omega)}^{k_- (\omega)} d q q \int_{0}^{\pi} d  \theta
 \frac{q^2}{ 2E_{q} } 
\frac{ \delta ( \theta - \theta_0) }{ |Jq \sin \theta|}, 
\end{align} 
where $\theta_0$ satisfies $\cos \theta_0 = (\omega - E_q )/(J q) $. 
As a result, we obtain 
\begin{align}
{\mathcal I}_{n} (\omega) 
= & 
 \frac{1}{4 \pi^2 c_1} 
\int_{k_+ }^{k_- } d q
 \frac{q^2  }{  \sqrt{ (f_+  - \omega)( \omega -  f_- ) } }, 
\end{align}
where we used 
$J q \sin \theta_0 = \sqrt{ (f_+  - \omega)( \omega -  f_- ) }$.  

When $\omega \ll \omega_{-}$, 
$k_{\pm } \simeq \omega/(c_{1} \pm J)$ holds. 
As a result, we obtain 
\begin{align}
{\mathcal I}_{n} (\omega) 
= & \frac{\omega^2}{4 \pi^2 c_1} 
\int_{ 1/(c_1 + J)^{}  }^{ 1/(c_1 - J)^{} }  
\frac{x^2 dx }{  \sqrt{ [(c_1 + J) x - 1] [ 1 - (c_1 - J) x ] } } 
\nonumber 
\\
= & 
\frac{\omega^2}{8 \pi c_1} 
\frac{2 c_{1}^{2} + J^{2}}{(c_{1}^{2} - J^{2})^{5/2}}. 
\label{newAppendixF21}
\end{align}

When $\omega_{-} \ll \omega \ll \omega_{+}$, 
the main contribution to the integral comes from $q \simeq k_- (\omega)$. 
In this case, we obtain $\omega \simeq E_{k_-} - J k_-$, and 
\begin{align}
f_+ - \omega \simeq  2 J k_- , 
\quad 
\omega - f_- \simeq \left . \frac{\partial f_-}{\partial q} \right |_{k_-} (k_- - q) . 
\end{align}
Introducing a proper cutoff $\Lambda = {\mathcal O}(\omega)$, 
and using 
\begin{align}
\left . \frac{\partial f_-}{\partial q}  \right |_{k_-} 
= \frac{\partial \omega}{\partial k_- }, 
\end{align}
we obtain 
\begin{align}
{\mathcal I}_{n} (\omega) 
\simeq &  \frac{1}{4 \pi^2 c_1} 
\int_{\Lambda}^{k_- } 
 \frac{q^2 dq }{  \sqrt{ 2 J k_-  }  \displaystyle{ 
 \sqrt{ 
\left .   \frac{\partial f_-}{\partial q} \right |_{k_-} (k_- - q) 
  } }  
  }
  \\
\simeq &  \frac{1}{4 \pi^2 c_1 \sqrt{2 J}} 
\int_{\Lambda}^{k_- } d q 
k_-^{3/2} \sqrt{ \frac{d k_-}{\partial \omega} } 
\frac{1}{\sqrt{k_- - q}}
\\
= &  \frac{1}{4 \pi^2 c_1 \sqrt{2 J}} 
k_-^{3/2}  \sqrt{ \frac{d k_- }{d \omega} } 
2 \sqrt{ k_-  - \Lambda }
\\
\simeq &  \frac{1}{4 \pi^2 c_1}
\sqrt { \frac{2}{J} } 
\sqrt{ \frac{d k_- }{d \omega} } 
k_-^{2} . 
\end{align}
Since $k_- \simeq (\omega / c_3)^{1/3}$, 
we end up with 
\begin{align}
{\mathcal I}_{n} (\omega) 
\simeq & 
 \frac{1}{4 \pi^2 c_1}
\sqrt { \frac{2}{3J} } 
\frac{\omega^{1/3}}{ c_3^{5/6}} . 
\label{newAppendixF28}
\end{align}

We evaluate the spectral function in the three-dimensional system, which is given by 
\begin{align}
{\mathcal I}_{n} (\omega) 
= & \frac{2 \pi}{(2\pi)^{3}} 
\int d q q^{2} \int_{-1}^{1} d (\cos \theta)  
\frac{q^2}{ 2E_{q} } 
\delta (\omega - E_{q} - J q \cos \theta) 
\nonumber 
\\ 
= & 
\frac{1}{(2\pi)^{2}} \int_{k_{+} }^{k_{-}} d q \frac{q^{2}}{2 c_{1} J }  
\\
= & 
\frac{1}{24 \pi^{2} J c_{1}} (k_{-}^{3} - k_{+}^{3}) . 
\end{align} 
When $\omega \ll \omega_{-}$, we find 
\begin{align}
{\mathcal I}_{n} (\omega) 
= & \frac{1}{24 \pi^{2} v c_{1}} \left [ \frac{\omega^{3}}{(c_{1}-J)^{3}} - \frac{\omega^{3}}{(c_{1}+J)^{3}}\right ]
\\
= & \frac{\omega^{3}}{12 \pi^{2} c_{1}} \frac{ 3c_{1}^{2} + J^{2} }{(c_{1}^{2} - J^{2})^{3}} . 
\label{newAppendixF33}
\end{align}
When $\omega_{-} \ll \omega \ll \omega_{+}$, we find 
\begin{align}
{\mathcal I}_{n} (\omega) 
\simeq & \frac{1}{24 \pi^{2} J c_{1}}  \frac{\omega^{}}{c_{3}}. 
\label{newAppendixF35}
\end{align}
 
We close this appendix with a summary of the low-energy behavior of 
the density spectral function within Bogoliubov theory in a uniform system. 
The concepts are totally different between the Feynman's single-mode approximation 
and the Bogoliubov approximation. 
However, if we set $c_1 = 1$ and $c_3 = 1/8$, 
the approximations are mathematically equivalent in the low-energy regime. 
In fact, we have $|G|^{2} = k/2$ and $\sqrt{ ( k^{2}/2 ) (k^{2}/2 + 2)} \simeq k + k^{3}/8$ in a low-energy regime. 
When the system is stable, $J < J_{\rm c} (= 1)$, 
we can take the low-energy such that $\omega \ll \omega_{-} = \sqrt{8 (1 - J)^{3}}$. 
In this case, according to (\ref{newAppendixF13}), (\ref{newAppendixF21}) and (\ref{newAppendixF33}), 
we end up with (\ref{SMAblwqcBogo}). 
At the critical current $J = J_{\rm c} (= 1)$, we obtain $\omega_{-} = 0$, so that we consider the case $\omega_{-} \ll \omega$. 
According to (\ref{newAppendixF15}), (\ref{newAppendixF28}), and (\ref{newAppendixF35}), 
we end up with (\ref{SMAeqqcBogo}).

\end{document}